\documentclass[twocolumn,superscriptaddress,floatfix,showpacs]{revtex4-2}
\usepackage{graphics,amssymb,amsmath,epsfig,color}
\usepackage{graphicx}

\newcommand{\be}{\begin{equation}} \newcommand{\ee}{\end{equation}}
\newcommand{\bea}{\begin{eqnarray}} \newcommand{\eea}{\end{eqnarray}}
\begin{document}

\title{High-Precision Simulations of the Parity Conserving Directed Percolation Universality Class in 1+1 Dimensions}
\author{Peter Grassberger} \affiliation{JSC, FZ J\"ulich, D-52425 J\"ulich, Germany}
         \affiliation{Max Planck Institute for the Physics of Complex Systems, N\"othnitzer Strasse 38, D-01187 Dresden, Germany}
\date{\today}

\begin{abstract}

	Next to the directed percolation (DP) universality class, parity conserving directed percolation (pcDP; also called parity
	conserving branching annihilating random walks, pcBARW) is the second-most important model with an absorbing state 
	transition. Its distinction from ordinary DP is that particle number is conserved modulo 2, which implies that there are 
	two distinct sectors in systems with a finite initial number of particles: Realizations with even and 
	odd particle numbers show different scaling behaviors, and systems in the odd sector cannot die. 
	An intriguing feature of pcDP it is that some of its critical 
	exponents seem to be very simple rational numbers. The most prominent is the one describing the average number of 
	particles (or active sites) in the even sector, which is asymptotically constant. In contrast, the dynamical critical 
	exponent (which is the same in both sectors) seems not close to any simple rational. Finally, the order parameter exponent 
	$\beta$ (which is also the same in both sectors) is, according to the most precise previous simulations, 
	rather close to 1, but incompatible with it. We present high statistics simulations which 
	clarify this situation, and which indicate several other intriguing properties of pcPD clusters. In particular, we find 
	that all exponents which were close to rationals are even closer, and $\beta = 1.000$ with the error in the next digit.

\end{abstract}

\maketitle

\section{Introduction}

The parity conserving directed percolation (pcDP) universality class \cite{GKV,grass89}, also called even-offspring or 
parity conserving branching-annihilating random walks (pcBARW) \cite{takayasu92,Jensen94,zhong95,Tauber}, 
directed Ising model \cite{Menyhard,Meny-Odor}, generalized Domany-Kinzel model \cite{hinrichsen97},
or $\mathbb{Z}_2$ symmetric directed percolation (DP2) \cite{Ha25}, was introduced in \cite{GKV} as a counter 
example to the Janssen-Grassberger \cite{Janssen,Grass82} conjecture according to which absorbing state transitions with a 
unique absorbing state are always in the directed percolation (DP) universality class, provided they are continuous
(`2nd-order') transitions. In order to see why pcDP must be distinct from DP, it is 
most interesting in one spatial dimension, where it allows for several physical interpretations. 

The most straight forward is as directed percolation \cite{Hinrichsen} (or, if time is continuous, the contact process \cite{Hinrichsen}) 
with the added constraint that the particle number is conserved modulo 2 \cite{GKV}. Due to this, any initial configuration with 
an odd number of particles on a lattice with periodic b.c. cannot die, which is sufficient to establish it as distinct from ordinary 
DP. Conservation of particle numbers modulo 2 in 1 spatial dimension holds indeed in any model with two symmetric ground states, 
if the `particles' are interpreted as domain walls. This was indeed the case for the two models studied in \cite{GKV}, which were 
stochastic cellular automata with $Z_2$ symmetries. The two absorbing states are in this realization periodic states $\ldots+-+-\dots$
and $\ldots-+-+\dots$, and the dynamics correspond to suitable combinations of spin flip and spin exchange 
(Kawasaki) moves \cite{Menyhard}. Notice that 
this analogy between particle and domain wall models breaks down in higher dimensions, where particles still have dimension 0
while domain walls have dimension $D-1$.

The different names given to pcDP reflect the fact that it was re-discovered repeatedly, being a relatively natural construct.
But maybe its most remarkable aspect is the fact that some -- not all -- of its critical exponents are very simple rational numbers.
Take, e.g., the exponent $\eta_2$, defined via the average number of active sites in simulations starting with an even number of
particles (=active sites) on an infinite lattice, $N(t) \sim t^{\eta_2}$. Simulations show that $\eta_2=0$, i.e. $N(t) = const$
with very high precision \cite{Park13,grass13a}. Usually, simple critical exponents indicate that there exists a solvable model
in this universality class (although there exists at least one case where a solvable model has an irrational exponent
\cite{nolin23}), but even extensive analytical studies \cite{mussawisade} have not given any hint for any solvable model.
Moreover, another exponent, the dynamical exponent $z$ (which is defined via correlations in a stationary state on an 
infinite lattice, and is thus the same in the two sectors with even and odd particle numbers) is definitely not a {\it simple} 
rational number \cite{Park13,grass13a}. Finally, there is the order parameter exponent $\beta$, also defined via correlations in 
stationary states and thus sector independent. It is estimated in the most precise simulations \cite {Park20} as $\beta = 1.020(5)$, 
which is suggestively close to $\beta=1$ but still 4 standard deviations away.

It is the main purpose of the present study to improve on this unsatisfactory situation by simulations with very high statistics, 
which might hopefully decide with reasonable safety whether the exponents in pcPD are simple or not, and find thereby any hint for a 
possible solvability of the model. For this purpose, we spent about 28 years of CPU time on work stations and laptops. We studied 
also several observables not analyzed in previous studies. We found some remarkable features and a possible hint why $\eta_2=0$, 
but no clear indication for solvability. On the other hand, we found clear evidence that $\beta=1$ exactly.

\section{The model and computational details}\label{model}

The very first simulations of the pcPD class used stochastic cellular automata \cite{GKV,grass89} with two states (0/1) per site. The 
absorbing states were periodic in space, and the `particles' were domain walls where the periodicity was broken. It was obviously
this somewhat counterintuitive feature (dictated by the fact that this led to the simplest algorithms with the most local interactions)
which prevented the widespread acceptance of the model, and which implied that the universality class was re-invented in 
\cite{takayasu92} and \cite{Menyhard}.

An early exact result by Sudbury \cite{sudbury90} was that, when treated as a contact process (i.e. with continuous time)
there was no phase transition in $D=1$, if branching created new particles in nearest neighbor sites and immediate annihilation happens
when two particles try to occupy the same site: The model is always subcritical, 
i.e. activity always dies out. This was later understood
in \cite{grass13a} by starting with discrete time (and space), and taking the limit where the time step tends to zero, together 
with all branching and hopping rates. While this limit is trivial in DP, in the sense that branching rates scale proportional 
to the time step, this is not the case in pcDP: There, the branching rate scales at the critical point differently from 
the time step, so that the branching rate per unit of time diverges as the time step tends to zero.

In view of Sudbury's result, some authors followed the contact process strategy but with 
four particles created instead of two in each branching \cite{takayasu92}, others used three states per lattice site
\cite{hinrichsen97,jensen97}, and still others made the annihilation non-local \cite{zhong95}.

In \cite{grass13a} it was pointed out that these complications are not needed if one uses discrete time, and two models 
(termed `A' and `B') were studied in detail (a similar algorithm was proposed in \cite{Ha25}). While updating is random sequential
in any algorithm with continuous time, and much CPU time is spent in memory access, updating in models A and B is parallel and thus 
much faster. It seems also faster than in the more complex models with discrete time, although we have not made systematic
studies. In the present paper we used algorithm `A' of \cite{grass13a}, with branching parameter fixed at $q=0.5$ and hopping 
parameter $p$ used as control parameter. This was mainly done because in this way $p_c$ was close to 1/2, which makes most efficient
use of the random number generator. For the latter we used xoroshiro128.

We made several types of simulations, the main ones being:

(1) Simulations starting with a single active site (`odd sector'). As we pointed out, activity can then never die. The main observables 
are then the average number $N_1(t)$ of active sites at time $t$, and their average squared distance $R_1^2(t)$ from the seed site (the 
subscript `1' refers here to the number of seed sites). We expect the scaling laws at large $t$ and at the critical point $p=p_c$

\be 
   N_1(t) \sim t^{\eta_1},\qquad R_1^2(t) \sim t^{2/z}.          \label{NR_1}
\ee

For $p\neq p_c$ one has the usual homogeneous function ansatz
\be
   N_1(p,t) \sim t^{\eta_1}\Phi[(p-p_c)t^{1/\nu_t}]                \label{N_Phi}
\ee
and an analogous ansatz for $R_1^2(t)$. Notice that the dynamical exponent $z$ and the correlation time exponent $\nu_t$ do 
not need a subscript `1', since they are defined via correlations in extended states with infinite numbers of active sites, 
in which there is no distinction between the two sectors.  Writing Eq.~(\ref{N_Phi}) instead of the more usual ansatz 
with $(p-p_c)t^{1/\nu_t}$ replaced by $|p-p_c|^{\nu_t}t$ has the advantage that $\Phi[x]$ is analytic and non-zero at 
$x=0$.

From  Eq.(\ref{NR_1}) we can get a first estimate of the fractal dimension $D_f$ of active sites at fixed large values 
of $t$, $D_{f,1} = \lim_{t\to\infty} \ln N_1(t)/\ln R_1^2(t)/2 = \eta_1 z$. Notice that the fractal dimension, being defined 
in principle via correlations in infinitely extended stationary states, should be the same in both sectors, 
$D_{f,1} = D_{f,2} = D_f$. Verifying this will be an important consistency check of our analysis.

(1a) Although activity cannot die completely in these simulations, it can be reduced to a single site, being 
pinched between leftmost and rightmost active sites when they become equal. Let us call the $i$-th time when this happens the 
`pinch time' $t_i$, and $\tau_i = t_{i+1}-t_i$ is the lag between successive pinches. In addition, $A_i$ is the number 
of active space-time points with $t_i \leq t < t_{i+1}$, i.e. the total activity between successive pinches. Notice that all 
$\tau_i$ are statistically independent, as are also all $A_i$. As far as $\tau_i$ and $A_i$ are concerned, the pcDP is a 
renewal process \cite{cox62}. 

It is expected that the distributions of $\tau$ and of $A$ are power laws at $p=p_c$,

\be
   Q_1(\tau) \equiv {\rm Prob}\{t_{i+1}-t_i \geq \tau\} \sim \tau^{-\delta_1},\qquad P_A(A) \sim A^{\gamma_1}
\ee
and are described by ansatzes analogous to Eq.(\ref{N_Phi}) for $p\neq p_c$. This is completely analogous to 
the scaling laws for durations and total activities of clusters in simulations starting with an even number of 
seeds (`even sector'). Notice that $Q_1(t)$ is also the probability that the first pinch occurred at some time $t'\geq t$. 

We also define $P_{\rm pinch}(t)$ as the probability that there is a pinch at time $t$, whether this is the first pinch 
or not. Following \cite{godreche-luck,godreche} we expect

\be
   P_{\rm pinch}(t) \sim t^{\delta_1-1}.
\ee
The observables $Q_1(\tau), P_A(A)$ and $P_{\rm pinch}(t)$ were measured only in part of the statistics, with the intent to verify 
the scaling laws but without the ambition to produce precise estimates of the exponents.

2) Simulations starting with two neighboring active sites or, to simplify the analysis, at positions $x = \pm 1$ (`even sector').
Observables and exponents are indexed here with subscript `2'. The main observables are, in addition to $N_2(t)$ and $R_2^2(t)$,
the survival probability $P_2(t)$. They are expected to satisfy
\be
   N_2(t) \sim t^{\eta_2},\quad R_2^2(t) \sim t^{2/z},\quad P_2(t) \sim t^{-\delta_2} 
\ee
at the critical point and by ansatzes analogous to Eq.(\ref{N_Phi}) for $p\neq p_c$.
These simulations gave us the best estimates for the exponents $\eta_2, z, \nu_t,$ and $\delta_2$. As in the odd sector, we can 
also get an estimate of the fractal dimension and verify that indeed $D_{f,1} = D_{f,2}$. As in the simulations in the odd sector,
we can also get the average number of active sites {\it per active run}
\be
   M_i(t) \equiv N_i(t) / P_i(t) \sim
     \begin{cases}
	     t^{\eta_1}                & {\rm for}\;\; i = 1\\
	     t^{\eta_2 + \delta_2}     & {\rm for}\;\; i = 2 .
     \end{cases}
\ee
Although we do not have a rigorous argument, we expect that it becomes irrelevant at late time whether surviving clusters have an
even or odd number of active sites, and thus $M_1(t)$ and $M_2(t)$ scale with the same exponent,
\be
   \eta_1 = \eta_2 + \delta_2.                               \label{eta_delta}
\ee
In addition, we made also some 
similar / additional measurements with lower statistics, sufficient to verify the scaling laws but not improving the exponent 
estimates:

(2a) Instead of starting with two neighboring active sites, we started with 2 sites at $x = \pm r_0/2$. For times $t \ll r_0^z$,
the two clusters growing at positions $\pm r_0$ do not interact. Thus one has effectively two clusters following the odd sector
evolution, which cannot die. Thus $P_{2,r_0}(t)$ is expected to scale as 
\be
   P_{2,r_0}(t) \sim 
     \begin{cases}
	 1                  & \quad t \leq r_0\\ 
      (t/r_0^z)^{-\delta_2} & \quad t > r_0 .
     \end{cases}
     \label{Pr_0}
\ee
For large fixed $t$, we have thus the scaling $P_{2,r_0}(t) \sim r_0^{z \delta_2}$.

(2b) Consider the case where a simulation (starting with two neighboring seeds) survives for a long time and has, at large $t$,
exactly 2 surviving active sites a distance $d$ apart. There are two extreme situations how such an event could occur:\\
(i) there was only one small cluster at times immediately before $t$, and the cluster is shortly before dying;\\
(ii) the cluster did split up into two clusters with odd activities at an early time, and these clusters moved away from
each other. Each of them cannot die, but they can become pinched at the same time $t$.\\
In case (i), the two active sites will be near by, in case (ii) they will be far away. Which of them is more likely, i.e. 
what will be the distribution $P_d(d)$? Since both cases are reasonably likely, one should expect a 
broad distribution. Indeed we found a power law
\be
   P_d(d) \sim d^{-\alpha_2}
\ee
cut off at $d_{\rm max} \sim t^{1/z}$.

3) As we pointed out already, the fractal dimension $D_f$ of the activity at some fixed large time $t$ should be equal in 
both sectors, being defined for (quasi-)stationary states on infinite lattices. In order to measure this important quantity 
in an alternative way, we measured the activities at large times in odd size clusters on finite lattices of size $L$. We expect 
\be
   \lim_{t\to\infty} N_{1,L}(t) \sim L^{D_{f,1}}.
\ee 
In these simulations we measured also the average times between successive pinches, which should also satisfy a power 
law at criticality, 
\be
   \lim_{t\to\infty} \langle \tau_{1,L}\rangle \sim L^{\sigma_1}.
\ee

4) We made analogous runs also in the even sector, where we measured the average cluster life time (instead of the time between
successive pinches) as a function of $L$.

 In the last two simulations we did not find any surprises. Nor were the results sufficiently precise that we learned anything 
 new. In order to keep the length of the paper reasonable, we will not present more details.
 
5) In a last set of simulations we tried to measure the order parameter exponent $\beta$ which was in previous simulations the 
most difficult one to pin down precisely, due to very large corrections to scaling. For this we need stationary simulations on 
very large lattices and very close to the critical point. We used lattices up to size $L=2^{26}$. Previous authors reached 
stationary states by starting with fully occupied or half occupied lattices. After an initial transient, the density decays
at intermediate times as $\rho(t) \sim t^{-\delta}$, where $\delta = \delta_2$ according to \cite{mussawisade}. The stationary 
state is reached when $t^{-\delta} \approx (p_c-p)^\beta$, which will be extremely late when $p$ is close to $p_c$. Notice that 
starting from a state with low density and no correlations would not be of any use, since the density would first increase sharply, 
and the following approach to the stationary state would be essentially the same as for a fully occupied start. 

In order to reduce the total transient time by several orders of magnitude, we used a trick that 
was already used in \cite{oslo}: We start with a low density state, but with correlations similar to those in the stationary
state. To find these initial states, we use the fact that $\eta_2$ is very close to zero and, moreover,
corrections to the scaling law $N_2(t) \sim t^{\eta_2} = const$ are very small for $p\approx p_c$. If we thus start, on a lattice
of size $L$, with 
\be
   N_0 = (p-p_c)L/2                    \label{N_0}
\ee
randomly located nearest-neighbor pairs of active sites, the evolution will start for $p\approx p_c$ 
like that of $N_0/2$ independent clusters, and thus the number of active sites will be constant. Actually we found 
that the stationary state is reached most quickly when $N_0$ is slightly different from Eq.(\ref{N_0}).

\section{Numerical Analysis and Results}

\subsection{General Remarks}
Scaling laws like those formulated in the previous section always are supposed to hold only asymptotically. For finite $t$ or 
$L$, one has to expect corrections which typically are also power laws, but with negative exponents. There are two types of 
such finite size (or finite time) corrections: \\
(i) Analytic ones, resulting from the fact that e.g. the proper time variable 
in a scaling law like Eq.(\ref{NR_1}) should have been $t+1$ or $t-1$. Such corrections are typically of the form 
\be
   A(t) \sim t^\alpha (1+a_1/t^{\beta_1}+a_2/t^{\beta_2} \ldots)       \label{corr}
\ee
with integer values of the correction-to-scaling exponents $\beta_i$.

(ii) Non-analytic ones, resulting from the renormalization group flow. They are of the same form, but the $\beta_i$ are in general
non-integer. 

There can also be combinations of both, and the distinction between analytic and 
non-analytic exponents can be blurred if some $\beta_i$ is rational (as, e.g., in Figs.~\ref{fig-R1-t-corr},\ref{fig-R2-t-corr}), 
since together with $\beta_i$ also any multiple of it can be a correction exponent. Traditionally, one orders the correction terms
such that $0 < \beta_1 < \beta_2 < \ldots$.

Analogous corrections hold for scaling laws which do not apply when a variable (typically time $t$ or system size $L$) tends to
infinity, but when it tends to zero (i.e., when we approach a critical point. The only difference is that then the correction-to-scaling 
exponents are positive.


Missing knowledge of these corrections is the main problem in the analysis of critical phenomena. Notice that performing simple 
least-square fits, with complete neglect of any correction terms, would in general give very poor results. A popular 
method for taking into account at least the {\it leading} corrections to scaling, and used e.g. in \cite{zhong95,Park13}, 
is to use effective exponents defined via slopes in log-log plots at finite values of the scaling variable. This is very 
efficient, if there is only one term in Eq. (\ref{corr}). But there are documented cases \cite{grassberger-backbone,nolin23} 
were up to three correction terms were needed for acceptable fits, and using effective exponents easily can then lead to wrong results. 

In such cases also data collapse, a popular method for analyzing scaling laws like Eq.(\ref{N_Phi}), is bound to encounter 
problems. In the following we shall demonstrate these, but we shall also propose another method for analyzing single-variable
scaling laws such as Eq.(\ref{corr}): We just make explicit least-square fits with several (up to 10) 
correction terms.
The amplitudes $a_i$ of the higher terms obtained in this way will not be physically meaningful, but the 
leading correction terms will be. If the leading exponent $\alpha$ and the first sub-leading exponents 
$\beta_i$ are wrongly chosen, the amplitudes $a_i$ will sharply increase with $i$. This is because the leading terms 
will give wrong predictions for intermediate $t$, and the next term in the expansion has to correct them. But this term 
will give wrong results for even smaller $t$ and the next term has to correct these, etc. If, however, the leading 
terms (including the exponent of the next correction term) are correct, then the later terms can all be small. We shall present 
one spectacular case of such `benign' behavior in subsubsection \ref{order-par}.

\subsection{Value of $p_c$ and Exponents Exactly at the Critical Point}

 \begin{figure}
 \begin{center}
 \includegraphics[width=0.5\textwidth]{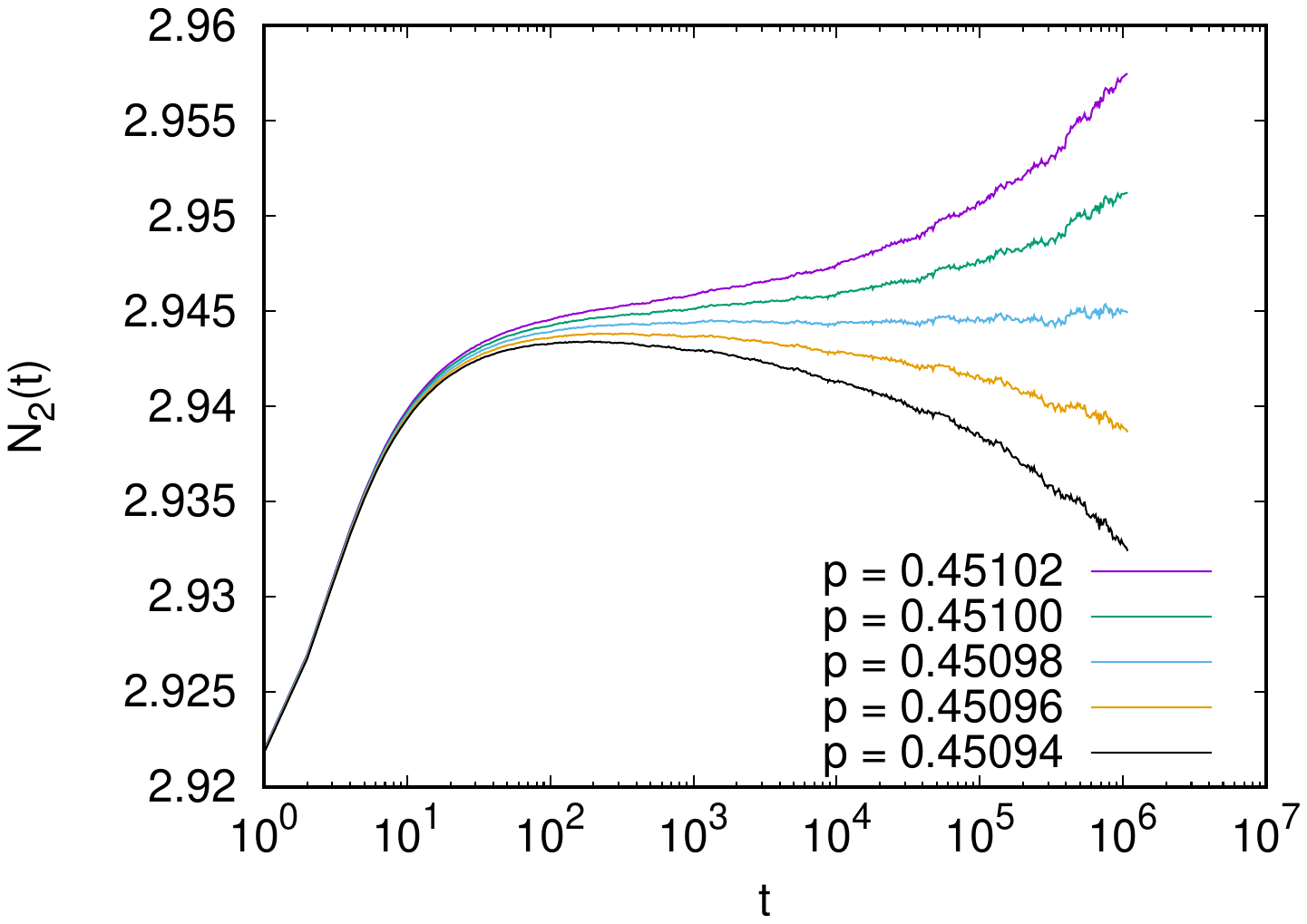}
	 \caption{(Color online) Log-linear plots of $N_2(t)$ against $t$, for five values of $p$ close to $p_c$. 
	 $N_2(t)$ is the average number of active sites in runs starting with two active sites at positions $x=\pm 1$.}
    \label{fig-N_2-t}
 \end{center}
 \end{figure}

In Fig.~\ref{fig-N_2-t} we show a log-linear plot of $N_2(t)$ versus $t$ for $0.45094 \leq p \leq 0.45102$. The critical point,
defined by a straight curve for large $t$, is obviously $p_c \approx 0.45098$ -- and the exponent $\eta_2$ is very close to zero.

 \begin{figure}
 \begin{center}
 \includegraphics[width=0.5\textwidth]{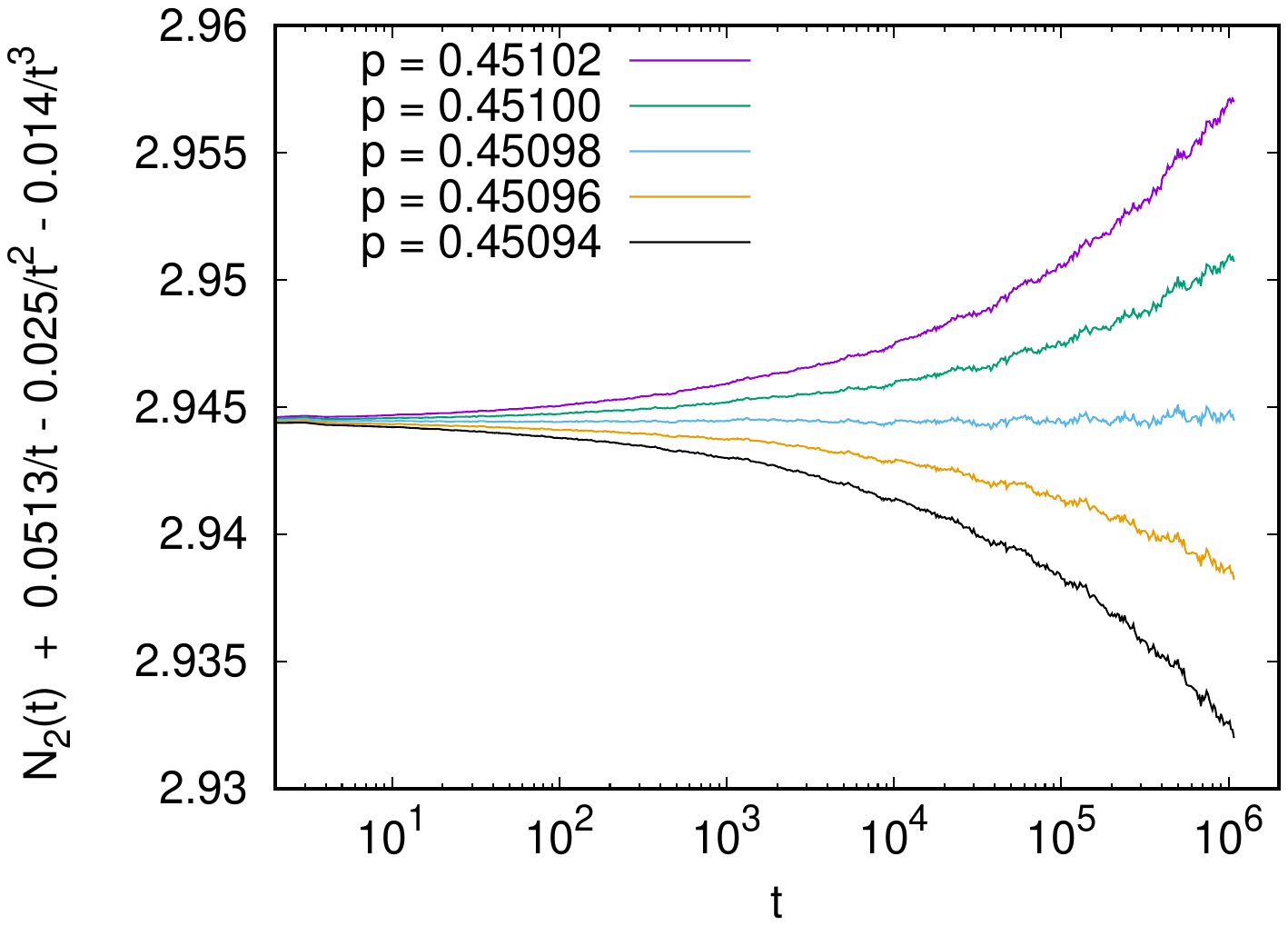}
	 \caption{(Color online) The same data as in the previous figure, but plotted with three correction terms with exponents 
	 -1, -2, and -3. Obviously these terms are sufficient to take into account all visible corrections to scaling.}
    \label{fig-N_2-t-corr}
 \end{center}
 \end{figure}

For more precise statements we have to deal more carefully with the corrections to scaling, which are obviously not negligible.
It is easy to see that the leading correction term has an exponent close to 1, but this term alone would not give an acceptable 
fit. In Fig.~\ref{fig-N_2-t-corr} we show a fit with three correction terms with exponents -1, -2, and -3. The amplitudes
in front of these exponential terms are small and {\it decrease with the rank}, suggesting thus that the fit is indeed
physically meaningful. Taking these scaling corrections into account, our estimates for $p_c$ and $\eta_2$ are
\be
   p_c = .450979(3) , \quad \eta_2 = 0.0000(1) .
\ee
This verifies the claim of \cite{Park13} that $\eta_2=0$ with extremely high precision.

 \begin{figure}
 \begin{center}
 \includegraphics[width=0.5\textwidth]{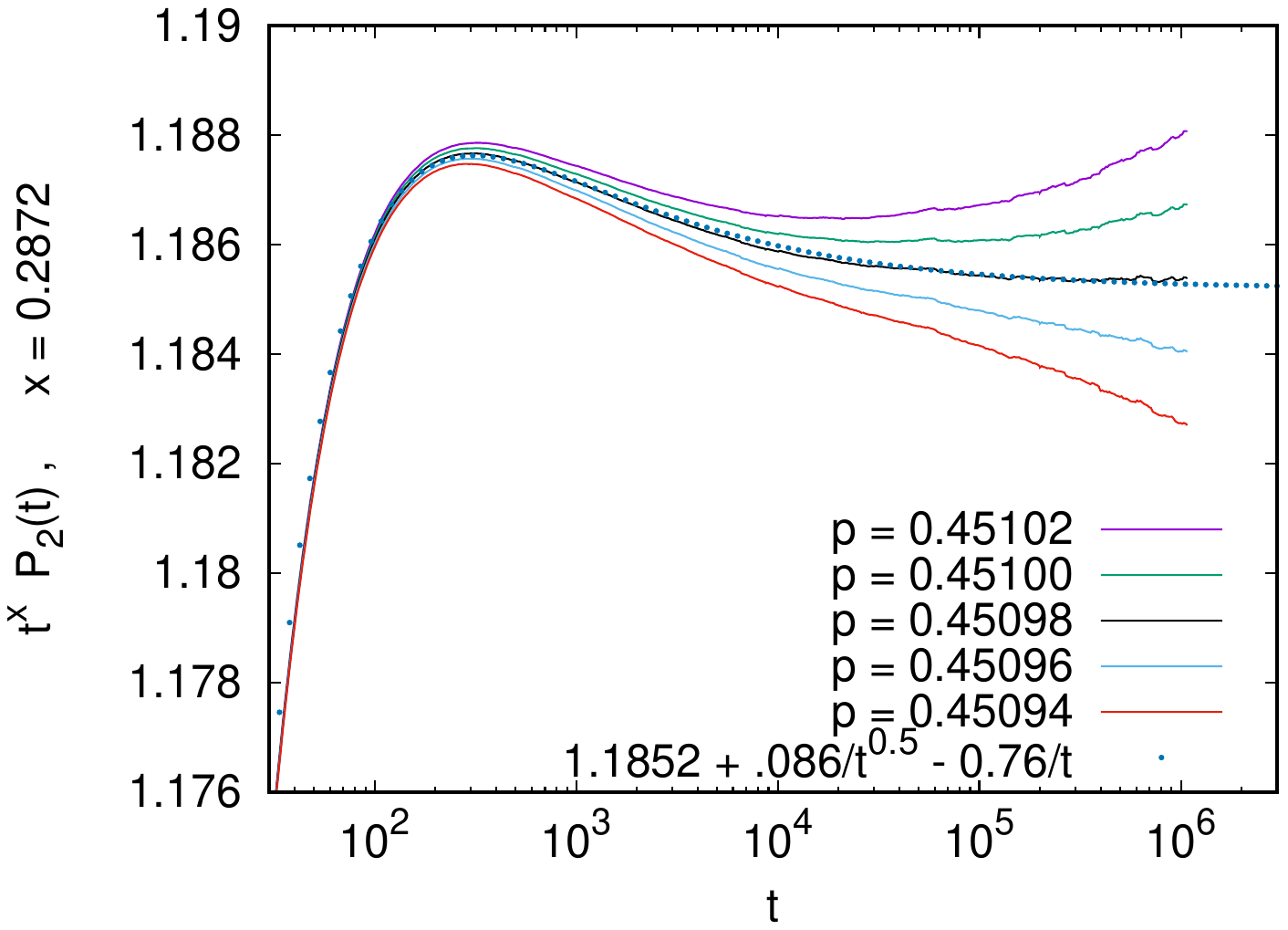}
	 \caption{(Color online) Log-linear plot of $t^{0.2872}P_2(t)$ versus $t$, where $0.2872$ is the value of $\delta_2$
	 estimated in \cite{Park13}. The dotted curve is an eyeball fit with a leading correction exponent $\beta_1=0.5$.}
    \label{fig-P_2-t}
 \end{center}
 \end{figure}

An analogous analysis of $P_2(t)$ was much less successful. In Fig.~\ref{fig-P_2-t} we show $t^x P_2(t)$, where $x=0.2872$
is the value of $\delta_2$ estimated in \cite{Park13}. In contrast to $N_2(t)$, the curve $p=p_c$ has now no definite curvature.
This means that definitely more than one correction term is needed for a quantitative fit. Also, the slow upward curvature for 
large $t$ suggests that the leading correction term has an exponent $\beta_1<1$ (roughly, we found $\beta_1 \approx 0.5$), 
i.e. there are non-analytic corrections while all correction in $N_2(t)$ were compatible with being analytic. But the most 
troubling fact is that we were not able to produce an acceptable fit, similar to Fig.~\ref{fig-N_2-t-corr}.

 \begin{figure}
 \begin{center}
 \includegraphics[width=0.5\textwidth]{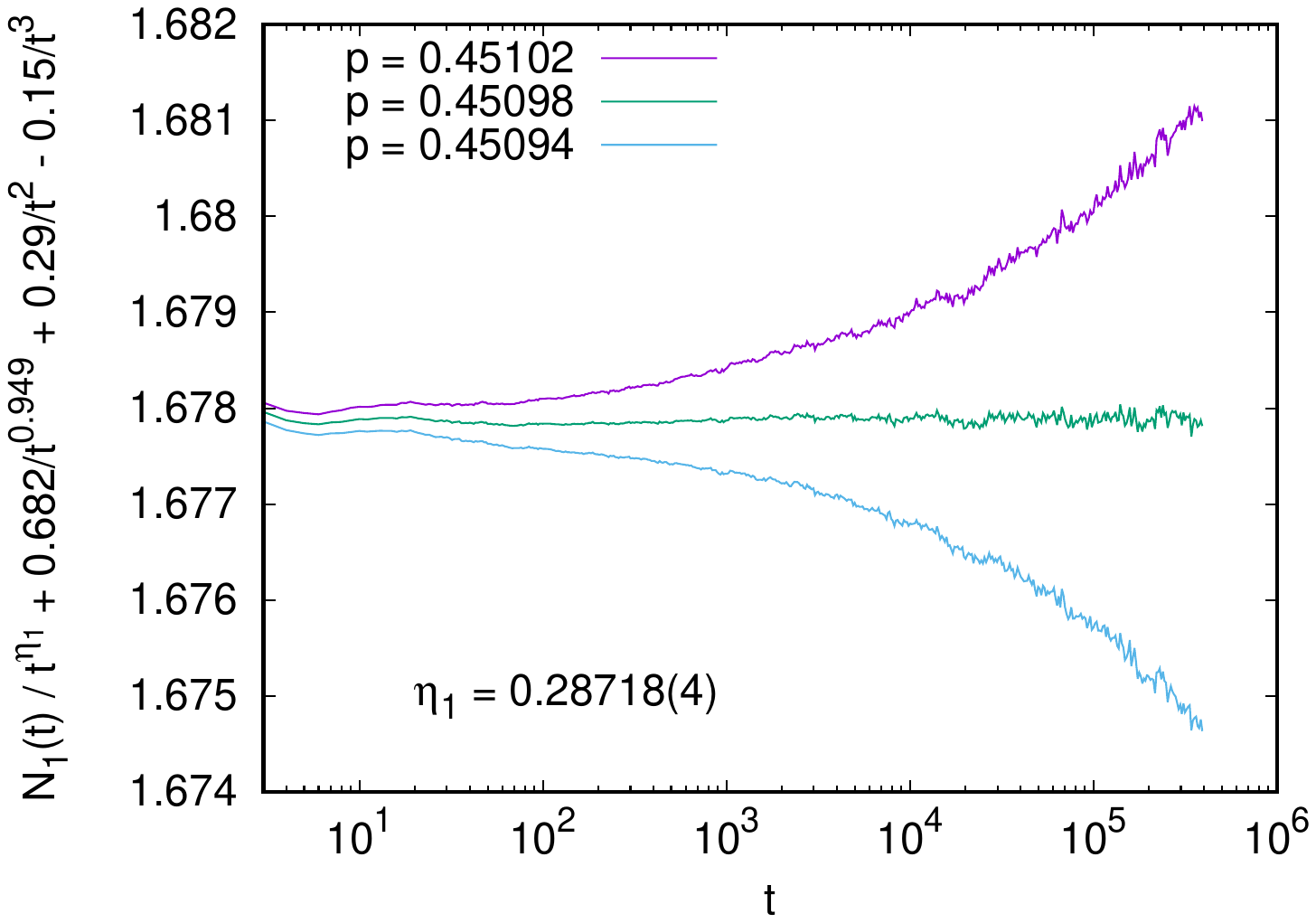}
         \caption{(Color online) Log-linear plot of $t^{0.28718}N_1(t)$ versus $t$, with three correction terms added so that
	 the central curve becomes flat for all $t$.}
    \label{fig-N1-t-corr}
 \end{center}
 \end{figure}

If $\eta_2=0$ as suggested by the data, then Eq.(\ref{eta_delta}) says that $\delta_2 = \eta_1$, so that we can estimate
$\delta_2$ also from simulations starting from a single active site. In Fig.~\ref{fig-N1-t-corr} we plot $t^{-\eta_1}N_1(t)$ 
against $t$, with three correction terms added. We see again a nearly perfect fit. It would suggest $p_c = .450981(5)$, which 
together with our previous estimate gives our preliminary  best value $p_c = .450979(3)$. Taking into account this uncertainty of 
$p_c$, our best estimate of $\eta_1$ is
\be
   \delta_2 = \eta_1 = 0.28719(5).        \label{delta}
\ee
This is four times more precise than the best previous estimate \cite{Park13}. It holds true (with doubled error), even if 
$\eta_2$ is not exactly zero.

In the fit shown in Fig.~\ref{fig-N1-t-corr}, the leading correction to scaling has an exponent -0.949. Taken at face value, this 
suggests a non-analytic correction term. But it also suggests that the leading term should actually have exponent -1 and is analytic.
We were able to produce an acceptable (although not as good as in  Fig.~\ref{fig-N1-t-corr}) fit with exponent -1, but only with
four correction terms and with amplitudes which increase rapidly with rank, so that the amplitude for the term with exponent -4
is $|a_4| > 10$. Thus the question is open whether the leading correction is non-analytic.

 \begin{figure}
 \begin{center}
 \includegraphics[width=0.5\textwidth]{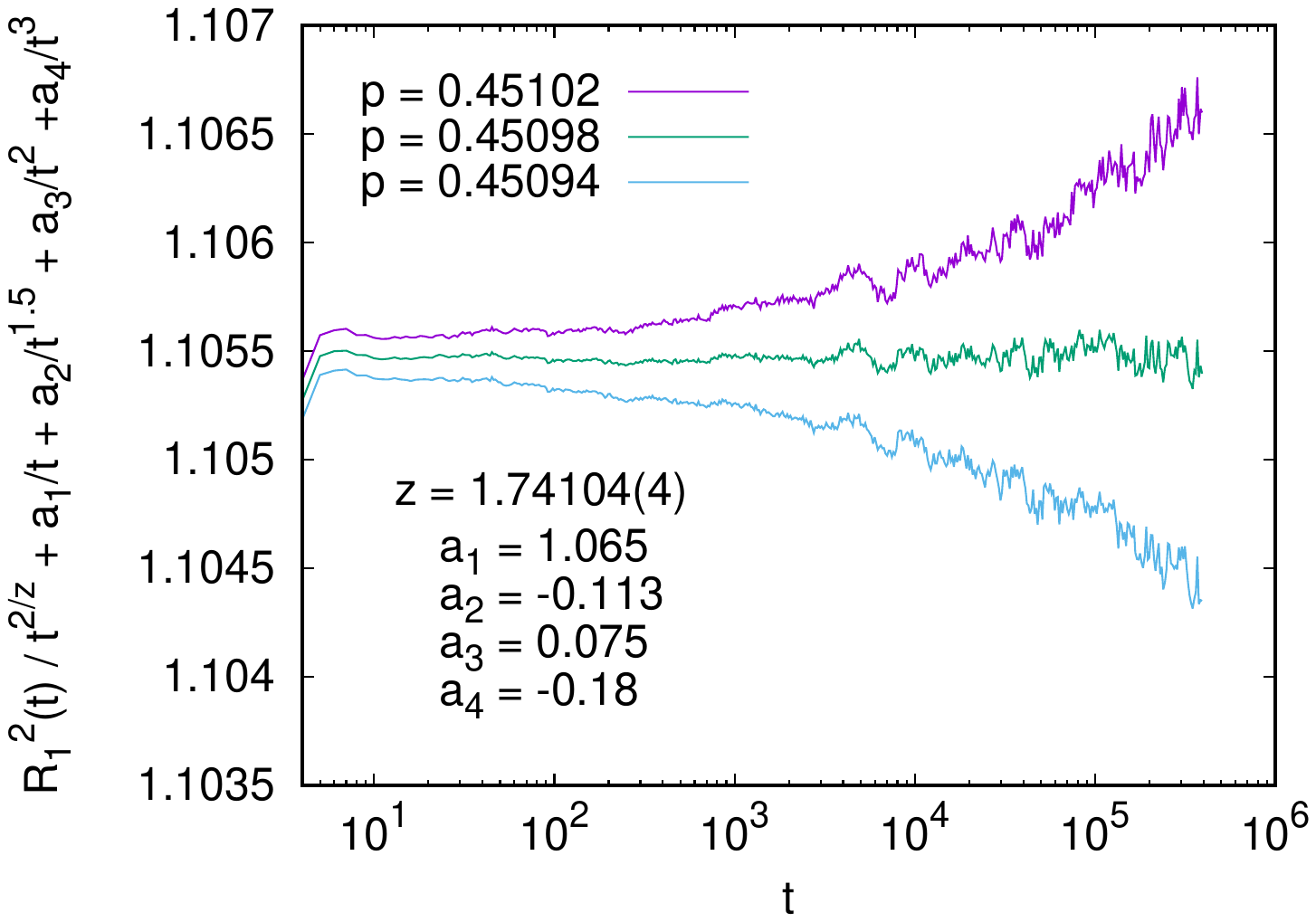}
         \caption{(Color online) Log-linear plot of $t^{2/z}R^2_1(t)$ versus $t$, with four correction terms added so that
         the central curve becomes flat for all $t$.}
    \label{fig-R1-t-corr}
 \end{center}
 \end{figure}

 \begin{figure}
 \begin{center}
 \includegraphics[width=0.5\textwidth]{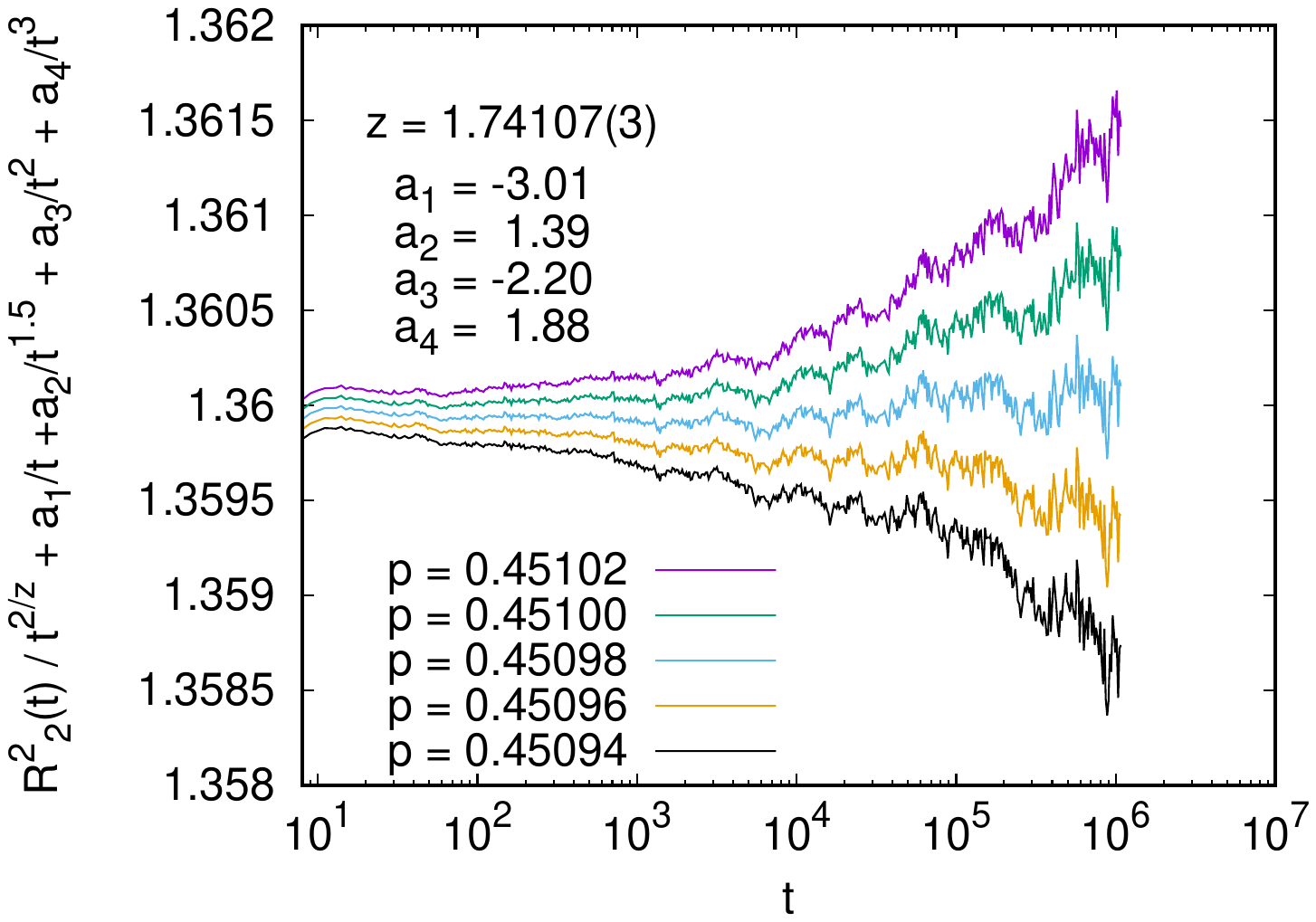}
	 \caption{(Color online) Same plot as the last one, but for $R_2(t)$ instead of $R_1(t)$.}
    \label{fig-R2-t-corr}
 \end{center}
 \end{figure}

The estimates for the dynamical exponent $z$ obtained from $R_1(t)$ and $R_2(t)$ are in perfect mutual agreement. The leading 
correction exponent is analytic with $\beta_1 = 1$. We show results for $R^2_1(t)$ in Fig.~\ref{fig-R1-t-corr}, and for $R^2_2(t)$ in 
Fig.~\ref{fig-R2-t-corr}. As the previous plots these are log-linear plots.
They show $R^2_i(t)/t^{2/z}$ (with $z$ being in each plot the best estimate) and with four correction terms, 
one of them being non-analytic with exponent -1.5 in both plots. The best estimate of $z$ combined from these two plots is 
\be
   z = 1.74106(3).                  \label{z}
\ee
It is about 13 times more precise than the best previous one \cite{Park13}. It is inconsistent with the conjecture $z=7/4$ of 
\cite{Jensen94}. It is however consistent with $z = 195/112 = 1.741071\ldots$, in which case we would have 
$\delta_2 = 56/195 =0.287179\ldots$.

The fact that $z$ is the same in the even and odd sectors, and that also the activity per surviving cluster scales with the same
exponent shows that the fractal dimension is the same too. From Eqs.~(\ref{delta}),~(\ref{z}) we obtain
\be
   D_f = \delta_2 z = 0.5000(1).
\ee
This suggests clearly that $D_f = 1/2$ exactly, as conjectured already in \cite{Jensen94,Park13}.

\subsection{Pinch Simulations in the Odd Sector}

 \begin{figure}
 \begin{center}
 \includegraphics[width=0.5\textwidth]{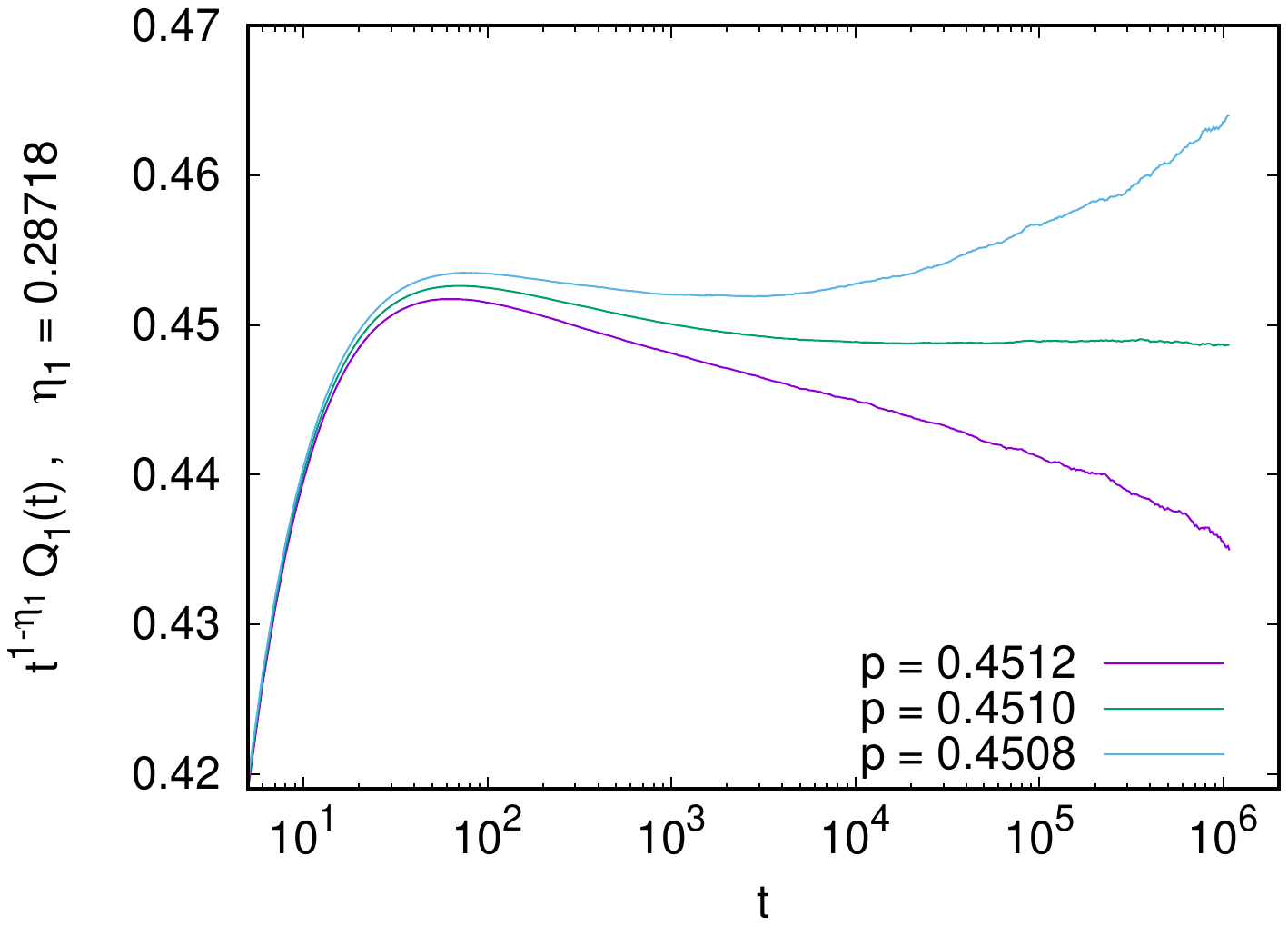}
	 \caption{(Color online) Log-linear plot of $t^{\delta_1} Q_1(t)$ versus $t$, with $\delta_1 = 1-\eta_1 = 0.71282$. }
    \label{fig-nopinch}
 \end{center}
 \end{figure}

In Fig.~\ref{fig-nopinch} we show data for the probability $Q_1(t)$ that the first pinch occurred at time $t' \geq t$, 
multiplied by a suitable power $t^{\delta_1}$ so that the curve for $p=p_c$ becomes flat asymptotically. We see that within 
errors
\be
   \delta_1 = 1-\eta_1.
\ee
As for $P_2$ (Fig.~\ref{fig-P_2-t}), the critical curve is not concave or convex, so that more than one correction to 
scaling terms would be needed, and a precise estimate of $\delta_1$ would be difficult (there is a slight downward curvature
in the curve for $p=0.4510$ for $t>3\times 10^5$, but this seems to be a statistical fluctuation).

We have no theoretical explanation why $\delta_1 = 1-\eta_1$, but the numerics seem clear. According to 
\cite{godreche-luck,godreche}, this implies that $P_{\rm pinch}(t)$, the probability that there is a pinch at time
$t$, should scale as 
\be
   P_{\rm pinch}(t) \sim t^{-\eta_1}.
\ee
We indeed verified this numerically (data not shown).

Finally, we verified that the r.m.s. radius $R^2(t)$ and the activity per `surviving'
event scaled, when conditioned on events with no pinch up to time $t$, as for the unconditioned sample.

\subsection{Pinches in the Even Sector}

\begin{figure}
 \begin{center}
 \includegraphics[width=0.5\textwidth]{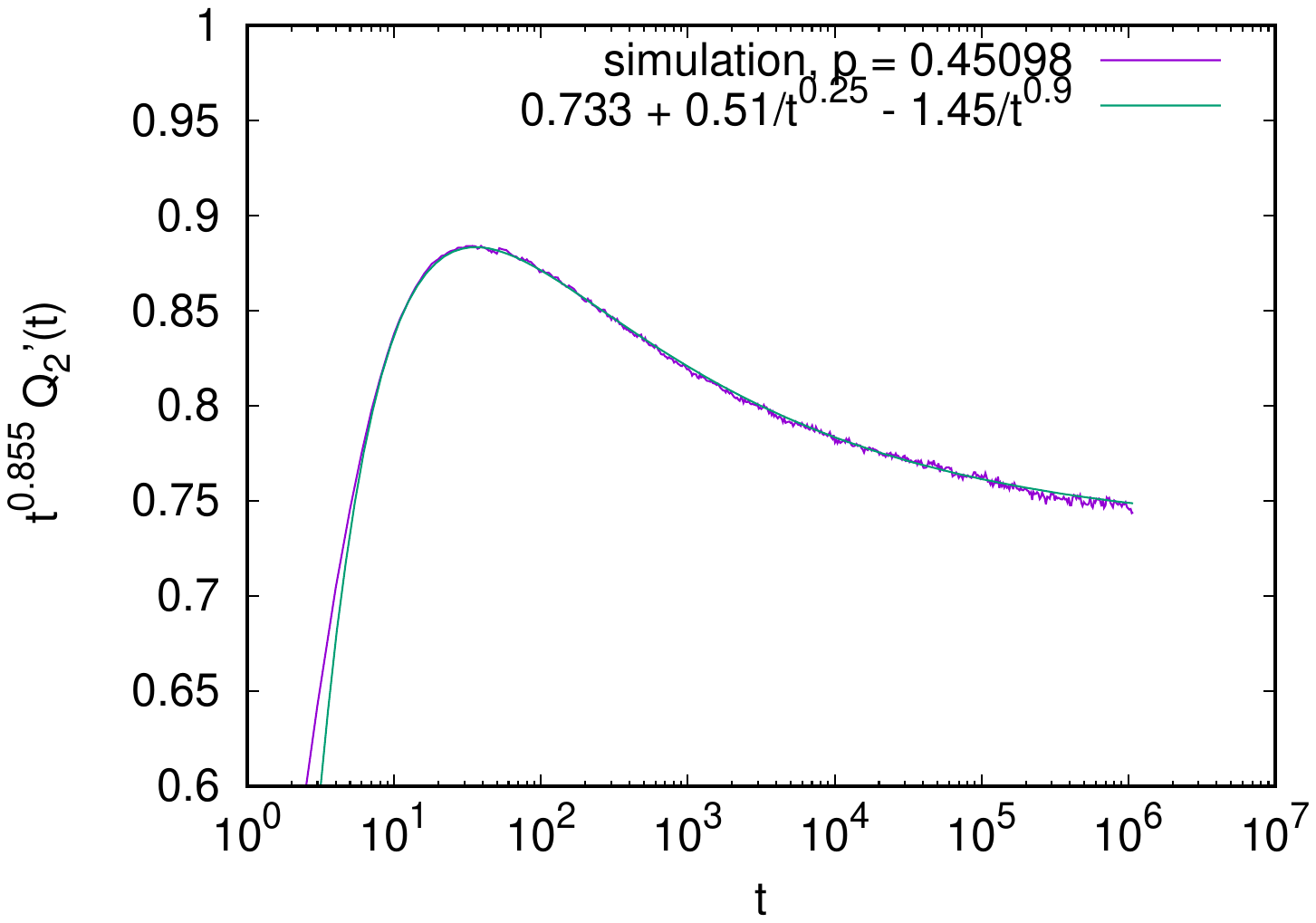}
	 \caption{(Color online) Log-linear plot of $Q_2'(t)$ versus $t$, the probability that there is a pinch at time $t$
	 in the even sector. In order to reduce statistical fluctuations, we actually
         lumped together data in narrow intervals $t'\in [0.99t,t]$. The smooth curve is a fit with two correction terms.}
	 \label{fig-Q_2'}
 \end{center}
 \end{figure}

In the even sector, `pinches' analogous to the above ones can be defined as events where the number of particles 
is reduced to 2. In a first set of simulations we measured the probability $Q_2(t)$ that the first pinch happened
at time $t' >t$, i.e. that there was no pinch at times $\leq t$. We found a clean power law with one important 
correction term with exponent -1/2 and with a new critical exponent, 
\be
   Q_2(t) \sim t^{-x} \times (1-0.865/\sqrt{t}-1.99/t) , \quad x = 1.0803(4).
\ee
Since pinches in the even sector do not, in contrast to pinches in the odd sector, imply a renewal structure,
the probability $Q'_2(t)$ that there is a pinch (not necessarily the first one) at time $t$ is not related to 
$Q_2(t)$, but follows an independent scaling law,
\be
   Q_2'(t) \sim t^{-y}, \quad y = 0.855(2).
\ee
As for $Q_2(t)$, we need again two correction terms (see Fig.~\ref{fig-Q_2'}), but the leading correction exponent is
$\approx 0.25$ instead of $\approx 0.5$. 
 
As we had discussed in Sec.~2, there are two extreme situation how a pinch in the even sector can occur, and as a result
we expect a power law interpolating between these two extremes.

 \begin{figure}
 \begin{center}
 \includegraphics[width=0.5\textwidth]{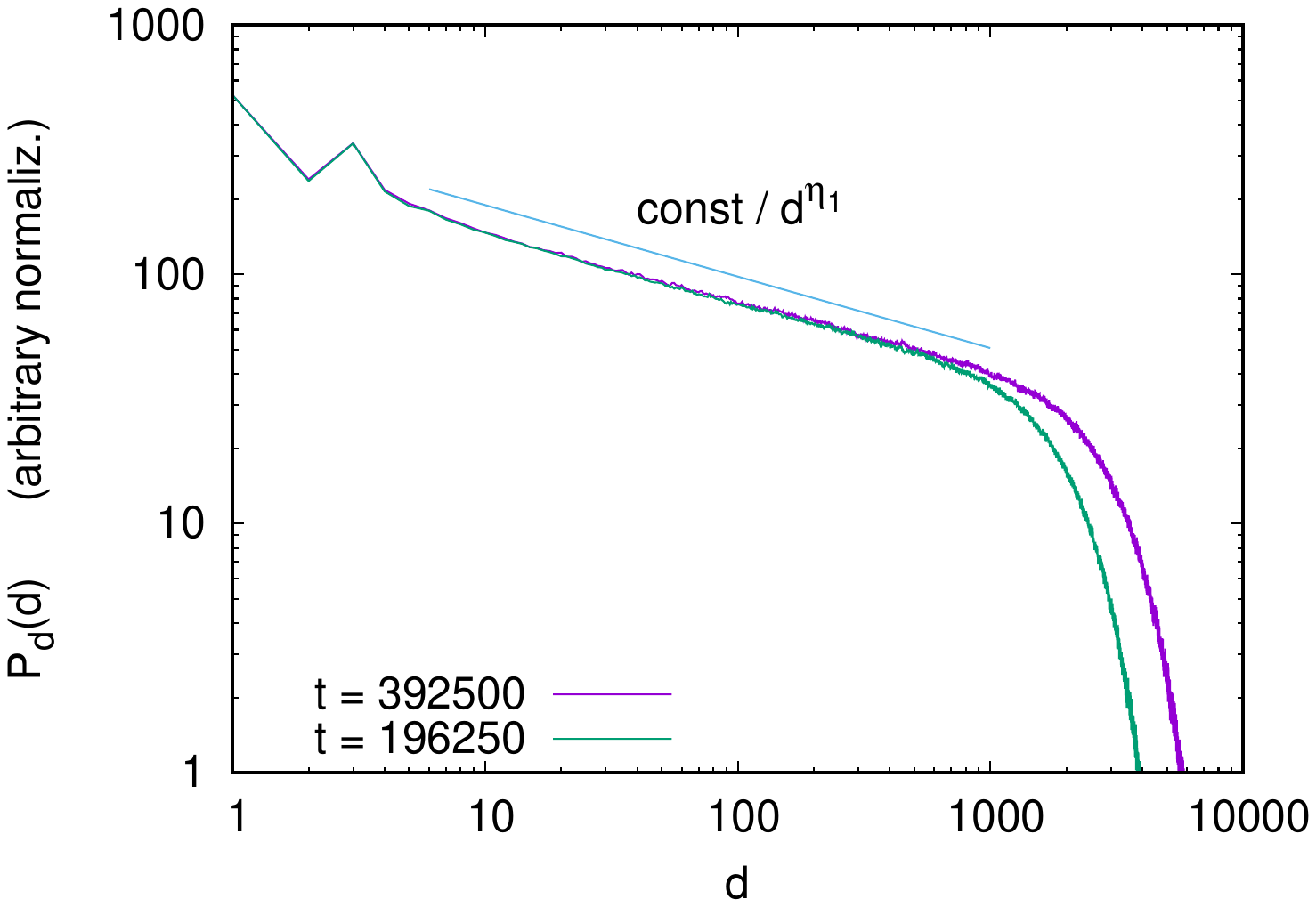}
	 \caption{(Color online) Log-log plot of $P_d(d)$ versus $d$, the distance between the two surviving particles 
	 in a quench in the even sector, for two values of $t$. In order to reduce statistical fluctuations, we actually
	 lumped together data in narrow intervals $t'\in [0.99t,t]$. The straight line is for a comparison with a power law
	 $P_d(d) \propto d^{-\eta_1}$.}
    \label{fig-even-pinch}
 \end{center}
 \end{figure}

In Fig.~\ref{fig-even-pinch} we show the (arbitrarily normalized) distributions of the distances $d$ between the 
two surviving particles, for $p=p_c$ and for two values of $t$. We see indeed a clear power law 
\be
   P_d(d) \sim d^{-\alpha_2},\quad \alpha_2 = 0.27(1)      \label{P_d}
\ee
with corrections at 
small $d$ and a sharp cut-off at large $d$. The exponent is suspiciously close to $-\eta_1$. Given the fact that two 
previous distributions (Figs.~\ref{fig-P_2-t} and \ref{fig-nopinch}) also had substantial corrections
which could have led to severe misestimations of the exponents, we believe that we cannot exclude the possibility
that actually $ \alpha_2 = \eta_1$, although it seems very unlikely in view of the following argument.

For $d=1$,  $P_d(d)$ scales with $t$ as the the probability that the event dies at time $t$, i.e. 
$P_d(d=1,t) \sim t^{-\delta_2}$. If Eq.~(\ref{P_d}) holds asymptotically in the entire range $1<d<t^{1/z}$, we 
obtain
\be
   y = 1-\frac{z-1}{2z^2}=0.8778(1),
\ee
which is close to our direct estimate $0.855(2)$, but incompatible with it.

\subsection{Even Sector with two Distant Point Seeds}

Finally, we made simulations with two active sites in the initial state, at positions $x = \pm r_0/2$, with 
$4 \leq r_0 \leq 256$. We verified that the resulting two clusters evolved independently (and could not die) up to a 
time $t_c \approx r_0^z$. For $t\gg t_c$, the survival probability scaled with $r_0$ as expected, see Eq.~(\ref{Pr_0}). 
We do not show any results, since they offered no surprises, and did not allow to estimate $z$ more precisely.

\subsection{The Correlation Length and Order Parameter Exponents}

From \cite{mussawisade} we know that the usual DP scaling relation $D_f = 1-\beta/\nu$ holds between fractal dimension, order 
parameter exponent $\beta$, and correlation length exponent $\nu$. Together with our result $D_f=1/2$ (at least 
within errors), this implies that 
\be
   \nu = 2\beta , \quad \nu_t = 2z\beta,        \label{nu_beta}
\ee
within very small errors.

In a last set of simulations we tried to estimate $\nu, \nu_t,$ and $\beta$ separately and directly. 
In measurements aimed at obtaining $\nu$ and $\nu_t$ we could clearly verify Eq.~(\ref{nu_beta}) within rather large errors, 
and we could reduce the errors quoted in \cite{Park13,Park20}. But success was limited, due to large and not well understood 
corrections to scaling. It was only when we measured the order parameter directly, that we could make a substantial break through.

\subsubsection{The Correlation Time Exponent} \label{corrlength}

The following analyses will only use the activity $N(t)$. 
In principle we could use also data for $P(t)$ and $R(t)$ to determine $\nu_t$, but they 
were always less informative. 

\begin{figure}
 \begin{center}
 \includegraphics[width=0.5\textwidth]{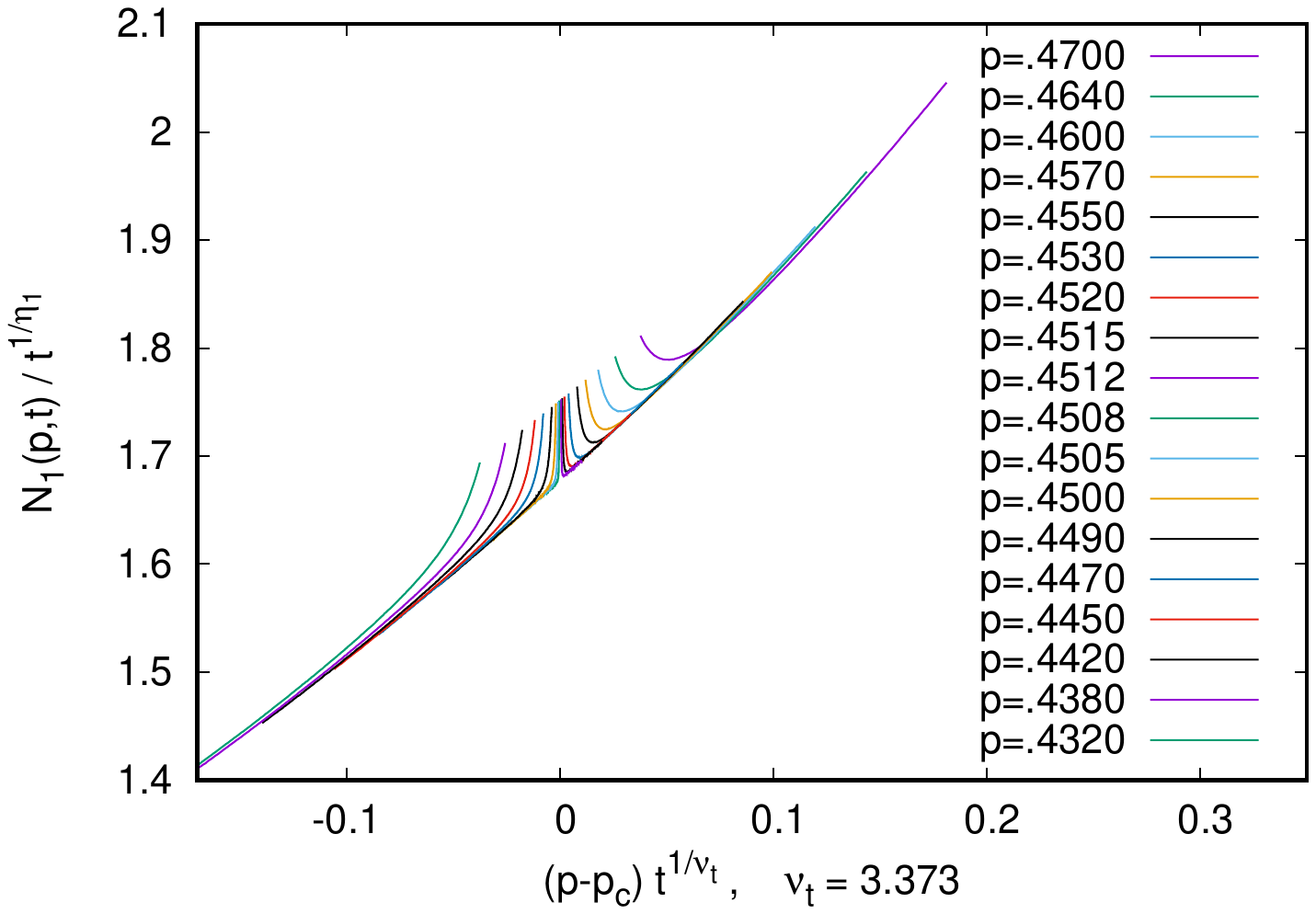}
	 \caption{(Color online) Linear-linear collapse plot of $N_1(p,t) / t^{\eta_1}$ versus $(p-p_c)t^{1/\nu_t}$, 
	 with $\nu_t = 3.373$ as a candidate value. The values of $p_c$ and $\eta_1$ are as determined above, the 
	 values of $p$ range from 0.432 to 0.47. Only values for $t\geq 10$ are shown.}
    \label{fig-single-collapse-0}
 \end{center}
 \end{figure}

{\bf Collapse plots for the odd sector:} As a first attempt to estimate $\nu_t$ we show in Fig.~\ref{fig-single-collapse-0} 
a collapse plot for the odd sector,
obtained by plotting $N_1(p,t) / t^{\eta_1}$ against $(p-p_c)t^{1/\nu_t}$, with $\nu_t = 3.373$ looking like the best 
compromise. The values of $p_c$ and $\eta_1$ are as determined above.

 \begin{figure}
 \begin{center}
 \includegraphics[width=0.5\textwidth]{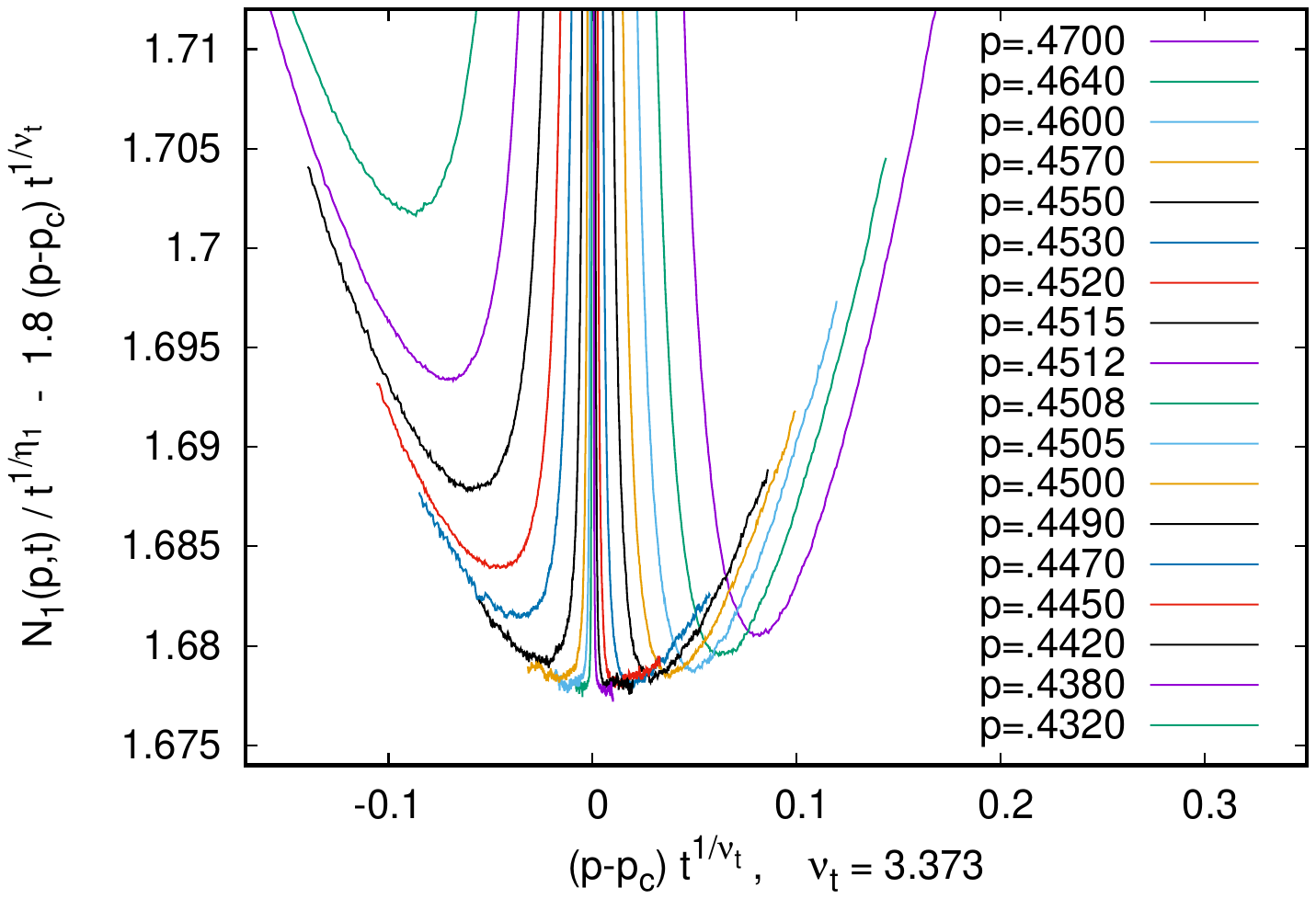}
         \caption{(Color online) Same data as in the previous plot, but with the x-axis tilted so that the
	 correlation between both axes is close to zero. The deviations from scaling for large values of $t$
	 and for $p$ not very close to $p_c$ are now clearly visible.}
    \label{fig-single-collapse}
 \end{center}
 \end{figure}

  \begin{figure}
 \begin{center}
 \includegraphics[width=0.5\textwidth]{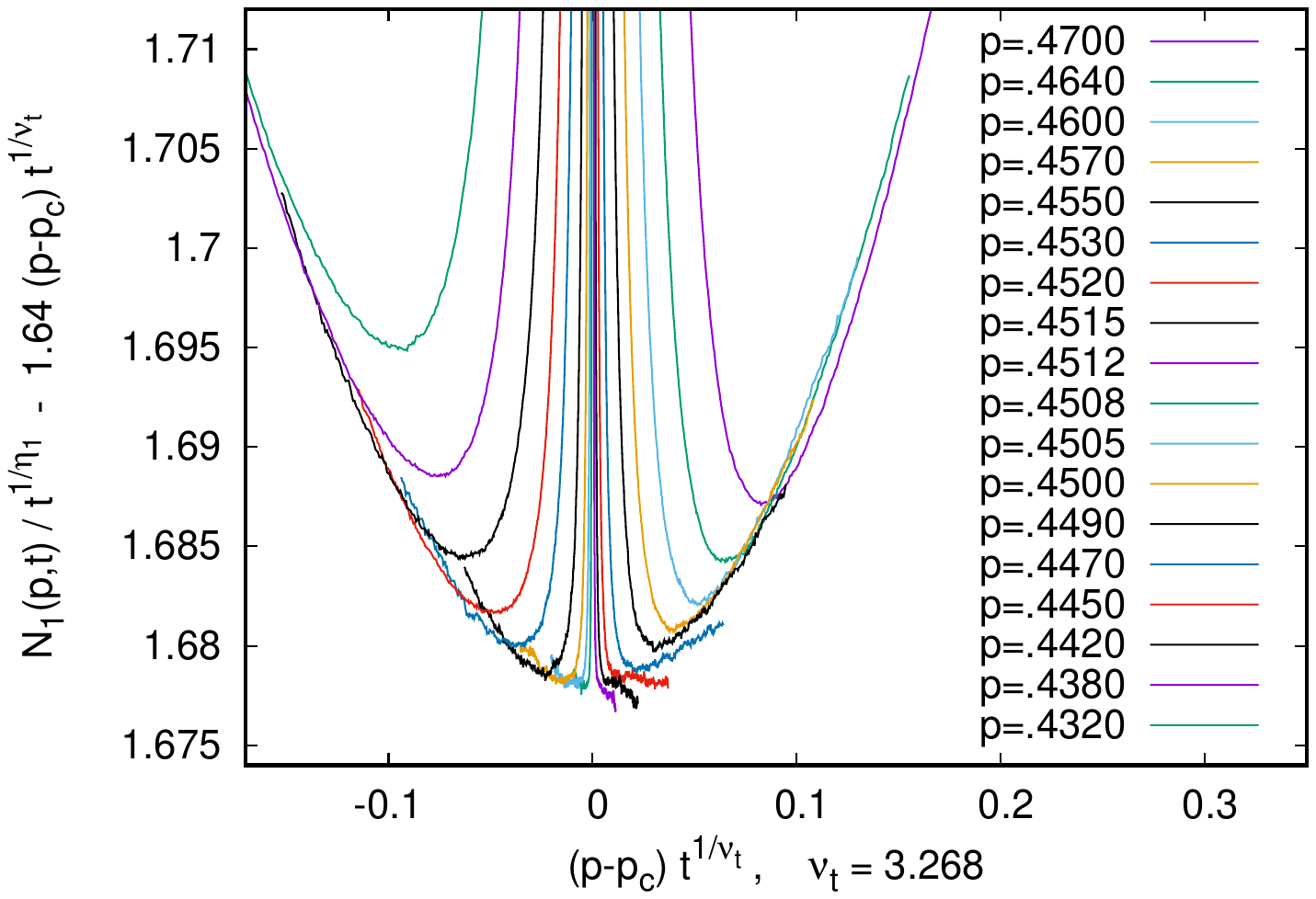}
         \caption{(Color online) Same data as in the previous plots, but with $\nu_t = 3.268$ instead of the 
	 value $\nu_t = 3.373$ used there. We see that there are now significant deviations at small $|p-p_c|$ and
	 large $t$, where the data collapse should be best.}
    \label{fig-single-collapse-alternative}
 \end{center}
 \end{figure}

We see a decent data collapse, with the main deviation at small $t$ (we plotted only values for $t\geq 10$
in this and the following plots; otherwise the deviations would look even worse). These deviations look ugly, but they 
are expected and not serious. A careful inspection suggest however that there might be smaller but systematic deviations 
for large $t$ and for $p$ not too close to $p_c$. In order to see these more clearly, plot the same data in 
Fig.~\ref{fig-single-collapse} after tilting the horizontal axis. Observe the much finer resolved y-axis in this plot.
It clearly shows that there are very significant scaling corrections, as soon as $p$ is not very close to $p_c$. One 
can get rid of these by changing the value of $\nu_t$, as done in Fig.~\ref{fig-single-collapse-alternative}, but then
there are significant deviations at $p\approx p_c$ and large $t$, where scaling should be best. We thus conclude 
that the alternative value $\nu_t = 3.268$ used in Fig.~\ref{fig-single-collapse-alternative} is too small, and our best 
estimate so far is $\nu_t = 3.37(5)$. 

 \begin{figure}
 \begin{center}
	 \includegraphics[width=0.5\textwidth]{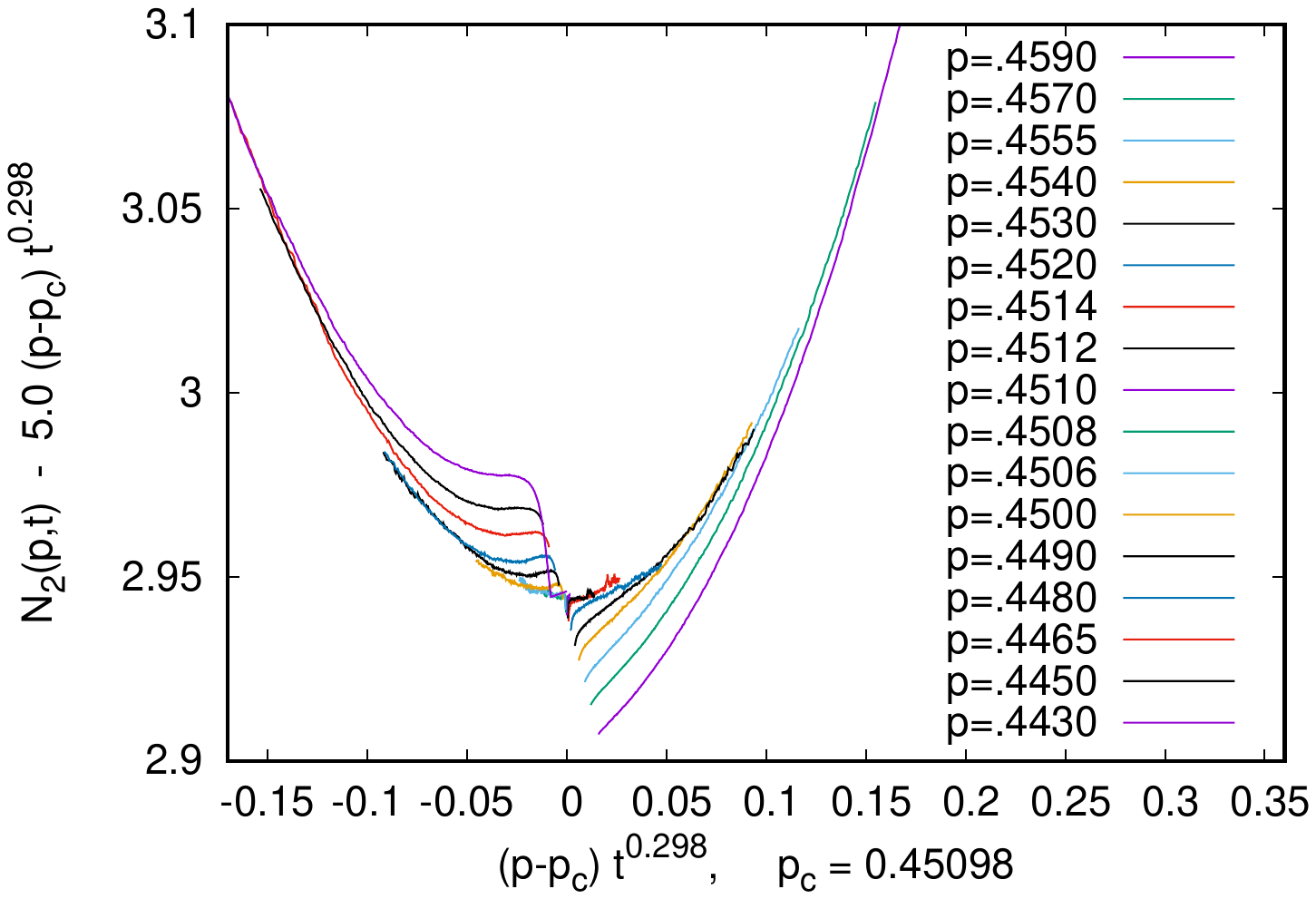}
	 \caption{(Color online) Collapse plot (with tilted x-axis) for $N_2(t)$, with $\nu_t = 3.356$. Again, only 
	 values for $t\geq 10$ are plotted.}
    \label{fig-N_2-collapse-alternative}
 \end{center}
 \end{figure}

{\bf Collapse plot for the even sector:} For the even sector, the situation is looks slightly better. In 
Fig.~\ref{fig-N_2-collapse-alternative} we show a collapse plot with tilted x-axis (analogous to 
Fig.~\ref{fig-single-collapse}) with $\nu_t = 3.356$. It shows very different (and strong!) scaling corrections for 
small $t$, but much better collapse at large $t$. On the other hand, similar plots for $3.31 < t < 3.41$ look also
acceptable. Thus we verify the estimate of $\nu_t$ obtained from the odd sector, but the errors reduce only marginally,
$\nu_t = 3.363(38)$ from both collapse analyses.

{\bf Taylor expansion for small $|p-p_c|$}

An alternative method for obtaining $\nu_t$ is to make a Taylor expansion in Eq.~(\ref{N_Phi}), which reads for the 
even sector
\be
   N_2(p,t) = t^{\eta_2}\{\Phi[0] +(p-p_c)t^{1/\nu_t}\Phi'[0] + \ldots\}.                \label{N_Phi_Taylor}
\ee
Thus
\bea
   \Delta(\epsilon,t) & \equiv & [N_2(p+\epsilon,t)-N_2(p-\epsilon,t)]/N_2(p,t) \nonumber \\
   & = & c \times \epsilon t^{1/\nu_t} + O(\epsilon^2),
\eea
and we can estimate $\nu_t$ from log-log plots of $\Delta(\epsilon,t)/\epsilon$ versus $t$ and extrapolating to $\epsilon 
\to 0$ (This is extremely efficient, if it is possible to obtain the derivative $\partial N/\partial \epsilon$ 
analytically \cite{grassberger-zhang} or if values of $N$ at $p, p+\epsilon$ and $p-\epsilon$ can be obtained from 
the same simulations \cite{ballesteros1997,dickman1999,newman2001fast,grass-highdim} so that statistical errors largely cancel, 
but we were not able to obtain either in the present case).

 \begin{figure}
 \begin{center}
	 \includegraphics[width=0.5\textwidth]{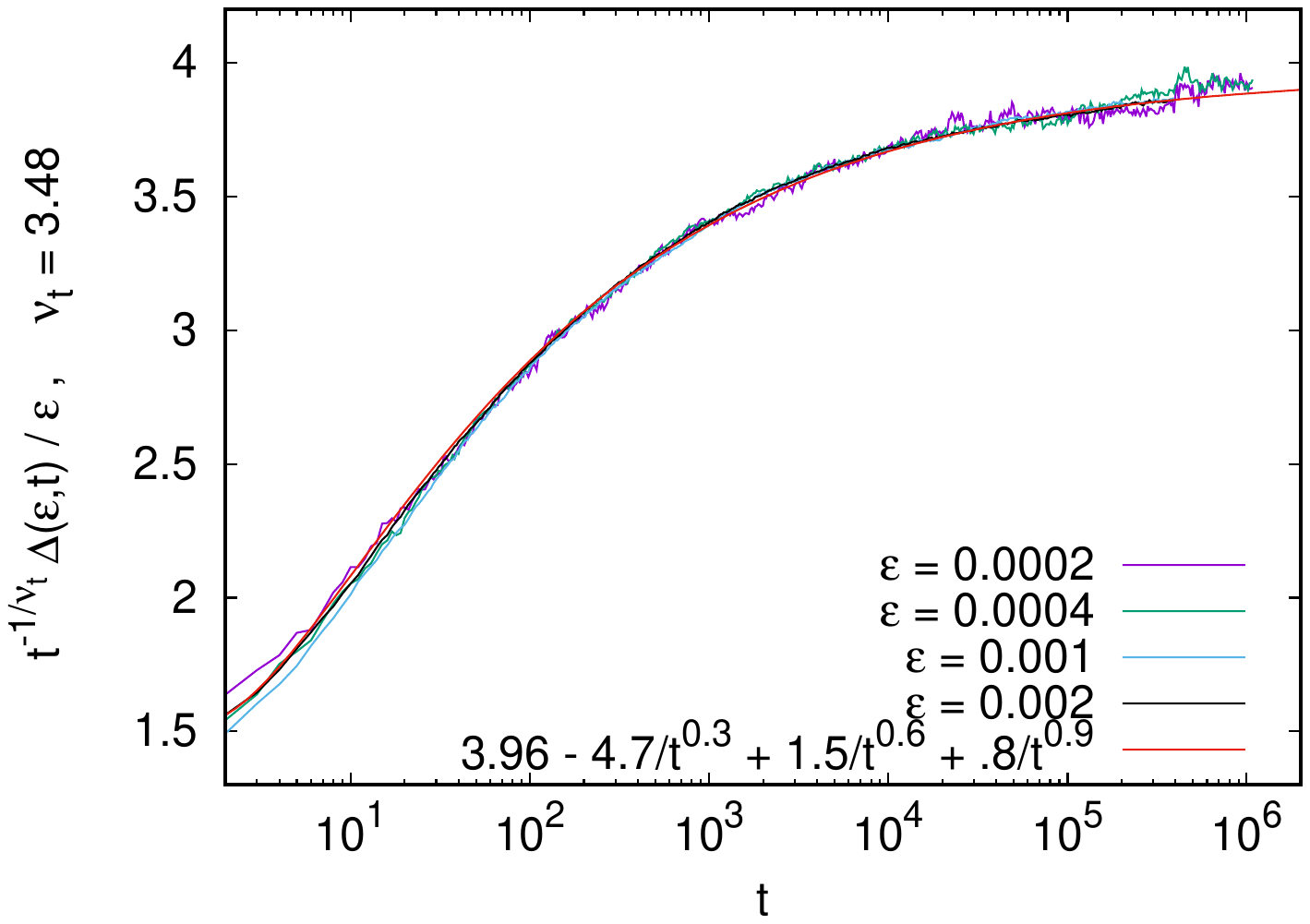}
	 \caption{(Color online) Log-linear plots of $t^{-0.2874} \Delta(\epsilon,t) / \epsilon $, for $\epsilon = 0.0002, 
	 0.0004, 0.001$ and $0.002$, in the even sector. The smooth curve is an eyeball fit with three correction to 
	 scaling terms, the leading one having exponent 0.3.}
    \label{fig-even-nu_t}
 \end{center}
 \end{figure}

Results of such a procedure for the even sector are shown in Fig.~\ref{fig-even-nu_t}. We used rather 
small values of $\epsilon$, to avoid the massive scaling violations for larger $\epsilon$ seen in 
Fig.~\ref{fig-N_2-collapse-alternative}. 
We see a perfect fit with three correction terms with exponents $\beta_1=0.3, \beta_2=2\beta_1,
\beta_3=3\beta_1$. But similarly good fits were obtained also within a rather large range of correction terms. Thus 
these data give only a rather uncertain estimate $\nu_t = 3.48(6)$. We do not show data for the odd sector, because 
they are much more noisy but consistent with this. 

Combining all three estimates of $\nu_t$ gives our preliminary best estimates
\be
   \nu_t = 3.39(3) ,\quad \nu = 1.95(2),  \quad \beta = 0.975(10),
\ee
where we also have used our previous values for $z$ and for $D_f$.

Our final estimates will take into account also the much more precise direct estimate for the order parameter $\beta$
obtained in the next subsection.

\subsubsection{The Order Parameter Exponent} \label{order-par}

In order to measure the order parameter exponent $\beta$ directly, we need estimates of the stationary density 
$\rho(p) = \lim_{t\to\infty} \rho(p,t) = \lim_{t\to\infty} N(t)/L$ 
at $p = p_c +\epsilon$, for very small values of $\epsilon$. Due to critical slowing down, simulations starting with 
finite initial densities take extremely long to reach stationarity. At the same time, in order to avoid finite size
corrections we need extremely large lattices.

We used lattices with $L$ up to $2^{26}$ and made $> 3\times 10^{9}$ time steps, for the smallest values of $\epsilon$. To 
reach these would have been impossible, if we had started with finite densities of order one. Instead, we used as 
initial states random configurations of next-nearest neighbor pairs with densities close to the expected stationary 
densities, as explained in Sec. 2 More 
precisely, using periodic boundary conditions we chose $N_0$ random pair positions $x_i,\; i=1,\ldots N_0$, and 
placed two particles at $(x_i\pm 1) \; {\rm mod}\; L$ for each $i$. If a site would become doubly occupied, one of the 
particles is discarded. The values chosen for $N_0$ were
\be
   N_0 = a \epsilon^{\beta_{\rm trial}},
\ee
where $a$ is a constant of order 1 (determined for each $\epsilon$ in some auxiliary runs), and $\beta_{\rm trial}$
is an estimated value of $\beta$. We used $\beta_{\rm trial}=1$.

\begin{figure}
 \begin{center}
	 \includegraphics[width=0.5\textwidth]{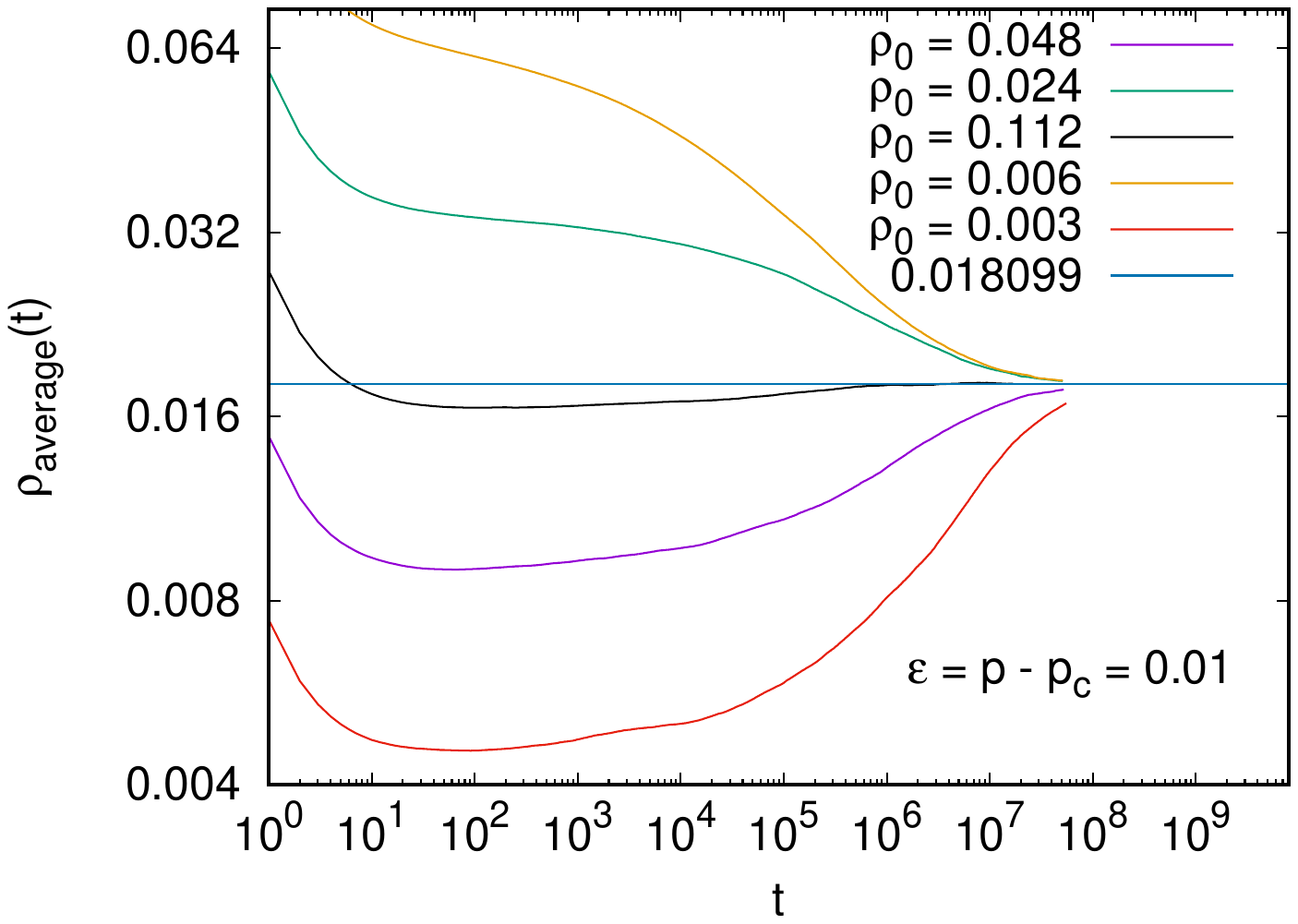}
	 \caption{(Color online) Time-averaged densities $\rho_{\rm average}(t) = t^{-1} \sum_{t' = 1}^t \rho(p,t')$
	 for $\epsilon = 0.01$ and five different values of $\rho_0 = 2N_0/L$. The central curve (for $\rho_0 = 0.0112$)
	 seems to show fastest convergence to the stationary value $\rho = 0.018099(8)$.}
	 \label{fig-rho-converge}
 \end{center}
 \end{figure}

The rationale behind this ansatz is the following: For small $\epsilon$, the pairs will be in average far apart from 
each other, and the evolution will start as with $N_0$ {\it independent} clusters. Since each cluster contains an
even number of active sites, it will have in average $\approx t^{\eta_2} \approx const$ active sites, and thus the 
density remains roughly constant until the clusters have grown sufficiently to touch each other. But by that time 
the correlations will have roughly reached stationarity, and the overall density remains roughly constant. This is 
illustrated in Fig.~\ref{fig-rho-converge} for $\epsilon = 0.01$. Notice that the main speed reduction is due to 
the fact that, with our algorithm, the CPU time needed for one time step is not $\propto L$, but $\propto N(t)$.

  \begin{figure}
 \begin{center}
         \includegraphics[width=0.5\textwidth]{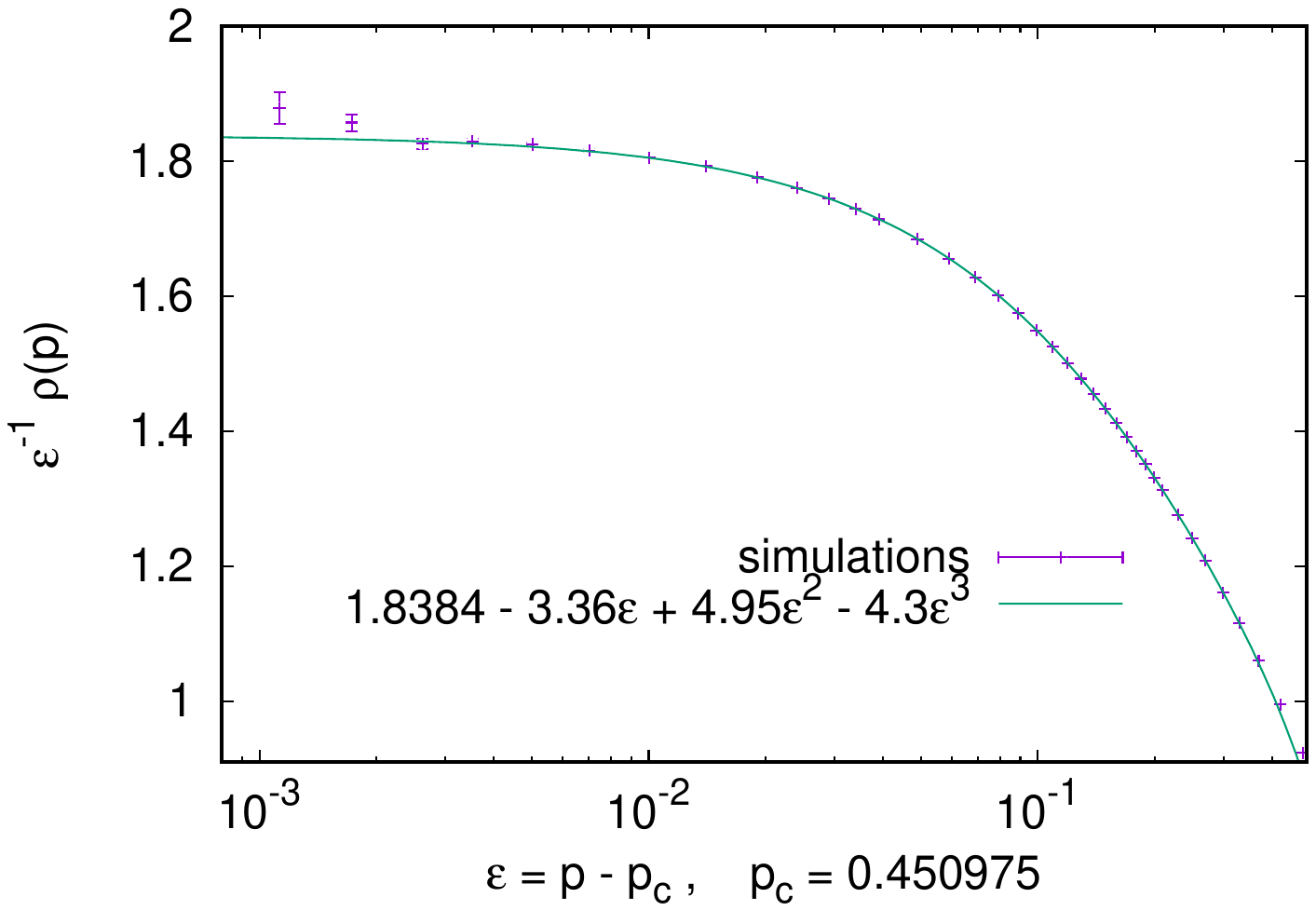}
         \caption{(Color online) Simulated values of $\rho(p)/\epsilon$ against $\epsilon$, together with a third order
	 polynomial eyeball fit.}
         \label{fig-density}
 \end{center}
 \end{figure}

Results of these simulations are shown in Fig.~\ref{fig-density}, where we plotted $\rho(p)/\epsilon$ against $\epsilon$.
Except for very small values of $\epsilon$ (where, however, statistical errors are large, and deviations are all smaller 
than $2\sigma$) the data are excellently fitted 
by a third order polynomial, the coefficients of which were just fitted by visual inspection. Obviously, Fig.~\ref{fig-density}
suggests that $\beta\approx 1$ with very high accuracy.

\begin{figure}
 \begin{center}
         \includegraphics[width=0.5\textwidth]{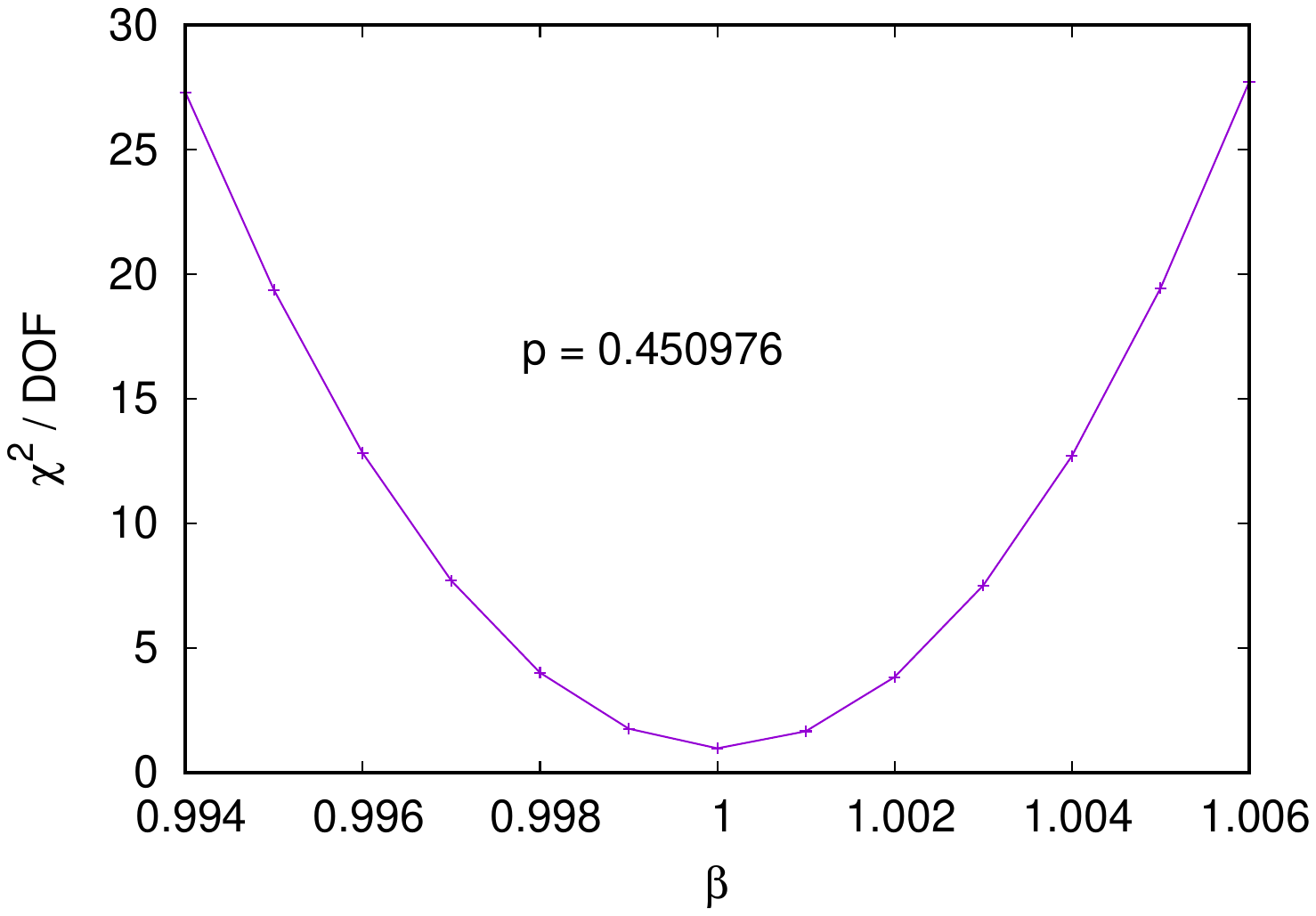}
         \caption{(Color online) Values of $\chi^2/DOF$, where $DOF = 38-11$ is the number of degrees of freedom in least 
	 square fits of $\rho(p)$ to a polynomials of 10th degree in $\epsilon$, for several values of $\beta$.}
         \label{fig-chi2}
 \end{center}
 \end{figure}

\begin{figure}
 \begin{center}
         \includegraphics[width=0.5\textwidth]{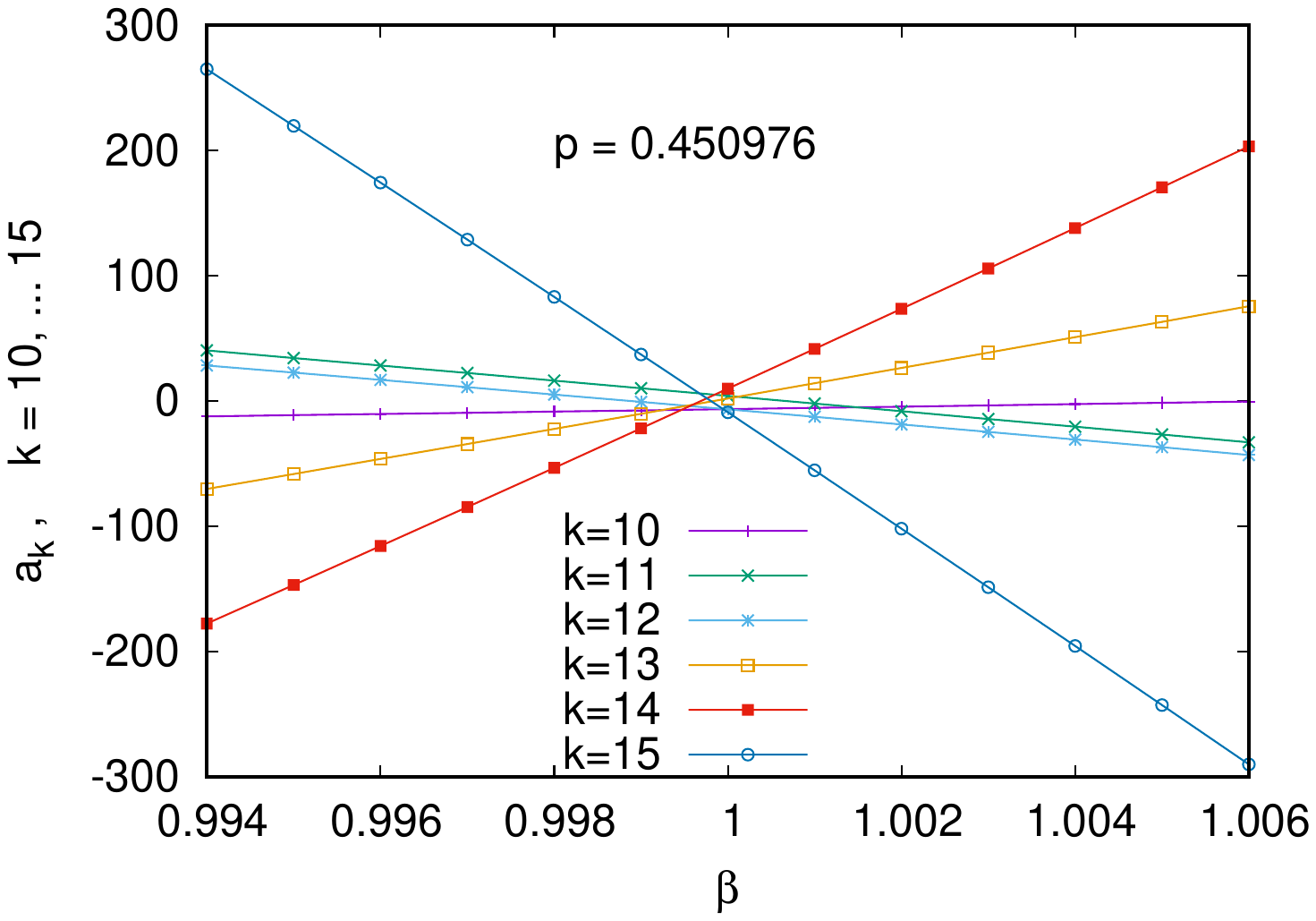}
         \caption{(Color online) The highest six coefficients in the least square fits,
         for the same values of $\beta$ as in the previous figure.}
         \label{fig-coeff}
 \end{center}
 \end{figure}

But in contrast to previous fits in this paper, we can now do much better, because the points in Fig.~\ref{fig-density} were
obtained in independent simulations and are thus statistically uncorrelated. We can thus try least square fits with much 
higher polynomials in $\epsilon$. Results from fitting polynomials of 10th order to $M=38$ data points 
$(\epsilon_i, \rho(p)/\epsilon_i^\beta; \; i = 1,\ldots 38))$ covering the range from $\epsilon_1 = 0.001125$ to 
$\epsilon_{38} = 0.93-p_c$, and with $p=p_c=0.450976$, are shown in Figs.~\ref{fig-chi2} and \ref{fig-coeff}. 

In Fig.~\ref{fig-chi2} we show the $\chi^2$ per degree of freedom (DOF) of these fits, as a function of $\beta$. If the fit is perfect,
we expect $\chi^2/DOF =1$ in the limit of large $M$. We see that $\chi^2$ has a minimum at $\beta = 1.0000(14)$ (where the error
bars take already into account the uncertainty of $p_c$) with $\chi^2/DOF =0.97$.

The six highest coefficients in the polynomials for the same value $p=0.450976$ are plotted against $\beta$ in Fig.~\ref{fig-coeff}. 
We see nearly perfect straight lines which cut the x-axis at $\beta = 0.9998(4)$. Taking into account the uncertainty of $p_c$ we 
obtain an error of $\beta$ comparable to that obtained from the minimum of $\chi^2$. The fact that all coefficients remain finite
at $\beta=1$ means that $\rho(p)$ is an analytic function of $p$ at $p=p_c$ within a circle of radius $\geq 1-p_c$.

We see thus that the direct measurement of $\beta$ gives an error which is {\it much} smaller than that obtained in the last 
subsection, whence we can nearly forget the latter and obtain the final estimates
\be
   \nu_t = 3.481(4) ,\quad \nu = 1.9994(22),  \quad \beta = 0.9997(11) .
\ee
Our obvious conjecture is that $\nu=2$ and $\beta=1$ exactly, and that $z=\nu_t/2=1/(2\delta_2)$ is the only exponent which is 
not a very simple rational.

Our final estimates for the critical exponents are summarized, together with the values obtained in \cite{Park20},
in Table 1. We see that we could in all cases reduce the statistical error. Moreover, in all cases where the exponents
are close to simple rationals, our estimates are closer to them that the estimates of \cite{Park20}. Our final estimate of 
the critical point is $p_c = 0.450977(2)$.

  \begin{table}
  \begin{center}
  \caption{Estimates of the critical exponents in Ref.~\cite{Park20} (middle column) and obtained in the
          present paper (r.h. column) }
  \begin{tabular}{c|c|c} \hline \hline
          $ \beta/\nu_{||}    $    & 0.2872(2) & 0.28719(5)   \\
          $ \beta/\nu_{\perp} $  & 0.5000(6) & 0.5000(1)     \\
          $ z = \nu_{||}/\nu_{\perp} $    & 1.7415(5) & 1.74104(4)\\
          $   \eta_1          $    & 0.0000(2) & 0.0000(1)\\
          $   \beta           $    & 1.020(5)  & 0.9997(11)\\
          $   \nu_{\perp}     $    & 2.04(1)   & 1.9994(22)\\
	  $   \nu_{||}        $    & 3.55(2)   & 3.481(4) \\ \hline\hline
\end{tabular}
\end{center}
\end{table}

\subsection{The Structure of Large Clusters: A Branching Process of Sub-clusters?}

  \begin{figure}
 \begin{center}
         \includegraphics[width=0.4\textwidth]{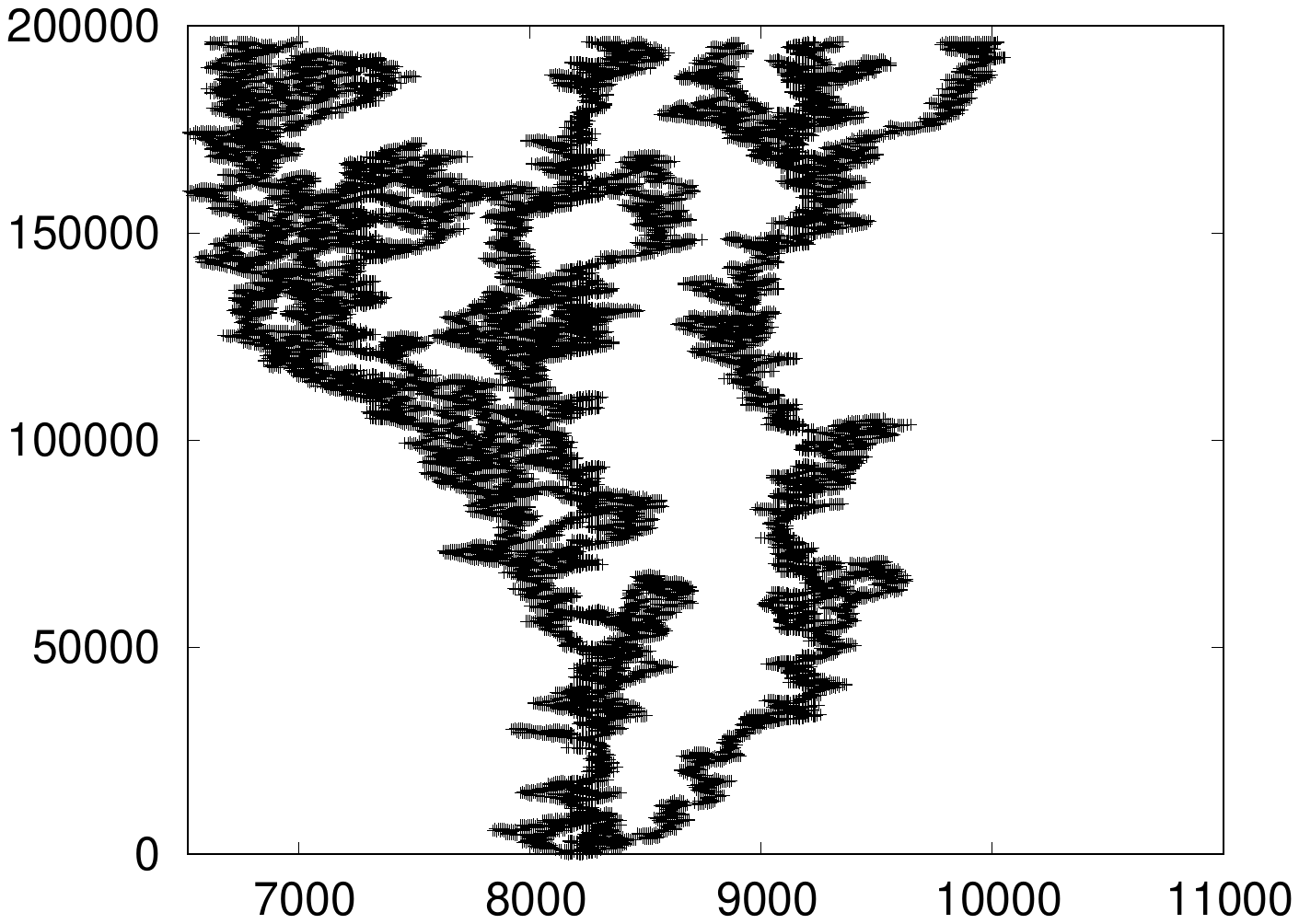}
         \includegraphics[width=0.4\textwidth]{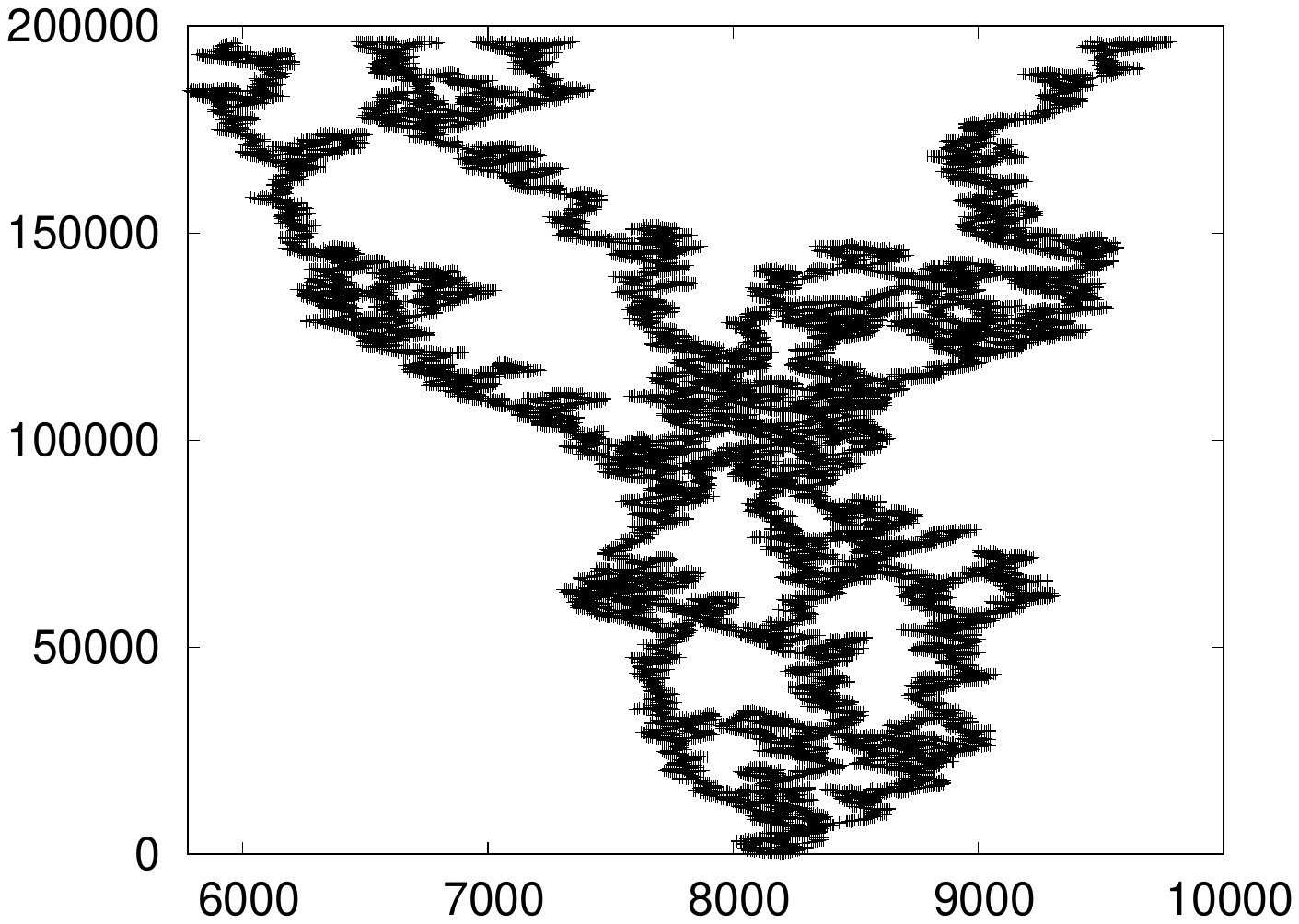}
	 \caption{(Color online) Typical clusters at $p=p_c$ in the (a) even and (b) odd sector. For large times, 
	 they consist of an increasing number of sub-clusters (`branches'), which touch each other less and less 
	 frequently as both time and the RG flow go on. Each sub-cluster is itself composed of sub-sub-clusters, etc.}
         \label{fig-picture}
 \end{center}
 \end{figure}

The success of the strategy for creating initial configurations shows also that active states at late times
consist essentially of randomly (but in average far away from each other) placed sub-clusters (branches), see 
Fig~\ref{fig-picture}. Since these 
sub-clusters can contain an even or odd number of particles, the number of sub-clusters is not necessarily even (odd) in 
the even (odd) sectors with odd particle numbers. 

For small $t$ these sub-clusters still touch frequently, but as time or coarse-graining increases the number 
of clusters increases and they become better and better defined, since they touch less and less frequently. 
In this limit, clusters show a clear hierarchical structure. Also, since sub-clusters
have longer and longer life times in this limit, they are more and more likely to have an odd particle number. Since the 
average number of sub-clusters also increases as $t\to\infty$, it is not clear whether clusters in the even 
(odd) sector clusters consist asymptotically of an even (odd) number of sub-clusters.

Anyhow, this picture may suggest that critical pcDP becomes asymptotically a branching process (in the strict sense) of 
sub-clusters. This would explain why $\eta=0$ in the even sector. It would however be in conflict with the fact that 
$\eta>0$ in the odd sector. It would also be in conflict with the fact that $z\neq 2$, i.e. that the sub-clusters 
diffuse anomalously.

\section{Discussion and Conclusions}

Although there are no known physical realizations of pcPD, this model is interesting for several reasons: First of all,
it is the prime example of an absorbing state transition which is not in the DP universality class. Apart from this,
it shows a remarkable set of unusual features:

(i) Some of the critical exponents are too close to simple rationals to be accidental. In particular, 
the exponent $\eta_2$ in the even sector is within very small errors compatible with being zero, and the fractal 
dimension of the active sites at fixed large times is $D_f = 1/2$. Finally, the order parameter exponent seems to 
be $\beta = 1$. For all of these three conjectures we gave additional evidence, based on vastly improved statistics.

(ii) In stark contrast to this, the dynamical exponent $z$ seems to be irrational. The rational candidate $z=r/q$ 
with the smallest denominator $q$ is $z=195/112$, which seems rather unlikely. We know of only one case (
2-d percolation and its backbone dimension \cite{nolin23}) where some exponents in the same universality class are 
rational and others not. The present simulations confirmed this situation by reducing the statistical uncertainty 
of $z$ considerably, to $z=1.74104(4)$.

(iii) The same irrational exponent (or, equivalently, the exponent $\delta_2 = 1/(2z) = 0.28718(5)$) seems to govern
two more, at first sight unrelated, observables:\\
(a) In the odd sector, the probability that a `pinch' occurs (where the number of active sites is reduced to 1)
at time $t$ decreases as $t^{-\delta_2}$; \\
(b) In the even sector, we can have similar pinches, where the number of active sites reduces to two. If the distance
between them is $d$, then we found that we have scaling for $1 \ll d \ll t^{1/z}$, with $P(d) \sim d^{-\delta_2}$. 

(iv) Different observables show vastly different corrections to scaling. While $N(t)$ and $R(t)$ in the even sector 
have corrections which are mostly or even entirely analytic (the leading correction exponent is 1 in both cases),
others like $P_2(t)$ have exponents $\beta_1 \approx 1/2$, and $Q_2'(t)$ (the probability, in the even sector, to have 
a `pinch' at time $t$)
seems to have even $\beta_1 = 1/4$. Moreover, in these cases there are usually more than two significant correction 
terms needed for quantitative fits, which makes their determination 
rather difficult. We found no evidence that the exponents of the leading correction terms were different in the 
even and odd sectors, but found that they differed for different observables. This might suggest that the leading
correction exponents are always the same, but coefficient
of of the leading correction vanishes for some observables. An exception is $N(t)$ in
the odd sector, where $\beta_1 = 0.949(5)$. This might suggest that the true exponent is $\beta_1 =1$, but the question is 
open.

Finally, we suggested a possible explanation why $\eta_2 = 0$ exactly. This would follow, if pcDP were a branching 
process in the strict sense (i.e., offsprings evolve independently), but this is obviously not the case {\it on the 
level of single particles}. We noticed, however, that the critical pcDP is approximately a branching process in 
which clusters split up into sub-clusters which interact very weakly at the critical point. Asymptotically (in the
RG sense) this may converge to a true branching process, at least in the even sector (sub-clusters are more 
crowded in the odd sector, whence the evolution in this sector stays more complex). If this 
scenario is more or less correct, it opens the fascinating possibility that the RG flow for some non-trivial
models can be from complex to structurally simple structures.

On the purely technical side we have observed that the order parameter seems to be an analytic function
of $\epsilon = p-p_c$ with a zero at $\epsilon = 0$.

In summary, we found that the pcDP universality class in 1+1 dimensions shows a number of very unusual features, 
some of which clearly suggest that a proper understanding can only (and should!) be obtained by analytic progress.

\section{Acknowledgements}

I am very much indebted to Jan Meinke for help with the computer systems at JSC.


\bibliography{mm}

\end{document}